\documentclass[twocolumn,showpacs,aps,floatfix,prd,nofootinbib]{revtex4}
\usepackage{graphicx}
\usepackage{dcolumn} 
\usepackage{amssymb} 
\usepackage{colordvi}
\usepackage{color}   
\usepackage{hhline}  
\usepackage{psfrag} 
\usepackage{epsfig}

\RequirePackage{xspace}

\newcommand{\BABARPubYear}    {13}
\newcommand{\BABARPubNumber}  {003}
\newcommand{\SLACPubNumber} {15524}
\newcommand{\LANLNumber} {1306.2895}

\input pubboard/babarsym

\long\def\inst#1{\par\nobreak\kern 4pt\nobreak
    {\it #1}\par\vskip 10pt plus 3pt minus 3pt}

\def\dEdx   {\ensuremath{{\rm d}E/{\rm d}x}\xspace}
\def\plab   {\ensuremath {p_{\rm lab}}\xspace} 
\def\pstar  {\ensuremath {p^*}\xspace} 
\def\thlab  {\ensuremath {\theta_{\rm lab}}\xspace} 
\def\thstr  {\ensuremath {\theta^*}\xspace} 
\def\cthlab {\ensuremath {\cos\thlab}\xspace} 
\def\cthstr {\ensuremath {\cos\thstr}\xspace} 
\def\cththr {\ensuremath {\cthstr_{\rm thrust}}\xspace} 
\def\etot   {\ensuremath {E_{\rm tot}}\xspace} 
\def\ththr  {\ensuremath {\theta^*_{\rm thrust}}\xspace} 
\def\cththr {\ensuremath {\cos\ththr}\xspace} 
\def\ecm    {\ensuremath {E_{\rm CM}}\xspace}
\def\mupm   {\ensuremath {\mu^\pm}\xspace}
\def\emp    {\ensuremath {(e\mu\pi)}\xspace}
\def\emppm  {\ensuremath {(e\mu\pi)^\pm}\xspace}

\def\eeqqb  {\ensuremath{\epem \!\!\to \! \qqbar}\xspace}

\def\pbar   {\ensuremath{\overline{p}}\xspace}
\def\ppbar  {\ensuremath{p/\overline{p}}\xspace}

\begin{document}

\preprint{\babar-PUB-\BABARPubYear/\BABARPubNumber} 
\preprint{SLAC-PUB-\SLACPubNumber} 

\begin{flushleft}
arXiv:\LANLNumber [hep-ex]\\
SLAC-PUB-\SLACPubNumber\\
\babar-PUB-\BABARPubYear/\BABARPubNumber\\
\end{flushleft}

\title{\boldmath
Production of charged pions, kaons and protons in \epem annihilations
into hadrons at $\sqrt{s} = 10.54~\gev$ 
}

%
\author{J.~P.~Lees}
\author{V.~Poireau}
\author{V.~Tisserand}
\affiliation{Laboratoire d'Annecy-le-Vieux de Physique des Particules (LAPP), Universit\'e de Savoie, CNRS/IN2P3,  F-74941 Annecy-Le-Vieux, France}
\author{E.~Grauges}
\affiliation{Universitat de Barcelona, Facultat de Fisica, Departament ECM, E-08028 Barcelona, Spain }
\author{A.~Palano$^{ab}$ }
\affiliation{INFN Sezione di Bari$^{a}$; Dipartimento di Fisica, Universit\`a di Bari$^{b}$, I-70126 Bari, Italy }
\author{G.~Eigen}
\author{B.~Stugu}
\affiliation{University of Bergen, Institute of Physics, N-5007 Bergen, Norway }
\author{D.~N.~Brown}
\author{L.~T.~Kerth}
\author{Yu.~G.~Kolomensky}
\author{M.~Lee
}
\author{G.~Lynch}
\affiliation{Lawrence Berkeley National Laboratory and University of California, Berkeley, California 94720, USA }
% ...
\collaboration{The \babar\ Collaboration}
\noaffiliation


\date{11 June, 2013}

\begin{abstract}
Inclusive production cross sections of \pipm, \Kpm and \ppbar
per hadronic \epem annihilation event are measured 
at a center-of-mass energy of 10.54~\gev,
using a relatively small sample of very high quality data from the 
\babar\ experiment at the PEP-II $B$-factory 
at the SLAC National Accelerator Laboratory.
The drift chamber and Cherenkov detector provide clean samples
of identified \pipm, \Kpm, and \ppbar over a wide range of momenta.
Since the center-of-mass energy is below the threshold to produce
a \BB pair, with \B a bottom-quark meson,
these data represent a pure \eeqqb sample with four quark flavors,
and are used to test QCD predictions and hadronization models.
Combined with measurements at other energies, 
in particular at the $Z^0$ resonance, 
they also provide precise constraints on the scaling properties of the
hadronization process over a wide energy range.
\end{abstract}


\maketitle

\section{Introduction}
\label{sec:intro}

The production of hadrons from energetic quarks and gluons in high-energy
collisions is well described by qualitative models,
but there are few quantitative theoretical predictions.
Detailed experimental information about hadron production allows
the confining property of the strong interaction to be probed.  
An empirical understanding of confinement is important to the
interpretation of much current and future high-energy data, 
in which the observable products of interactions and decays of heavy
particles, 
known and yet to be discovered, appear as jets of hadrons.
Measurements involving identified hadrons probe the influence on this 
process of hadron masses and quantum numbers such as strangeness,
baryon number, and spin.

The process $\eeqqb \to hadrons$ is understood to proceed through
three stages.
First, 
the quark ($q$) and antiquark (\qbar) ``fragment" via the radiation of
gluons ($g$),
each of which can radiate further gluons or split into a \qqbar pair.
This process is, in principle, calculable in perturbative quantum
chromodynamics (QCD),
and there are calculations for up to four final-state partons,
corresponding to second order in the strong coupling \as~\cite{ert},
where by ``parton" we mean either a quark or a gluon.
In addition, leading-order calculations exist for as many as six
partons~\cite{moretti},
as well as calculations to all orders in \as in the
modified leading logarithm approximation (MLLA)~\cite{mlla}.
There are also ``parton shower" Monte Carlo simulations~\cite{nlla} 
that include an arbitrary number of $q \!\to\! qg$, $g \!\to\! gg$ 
and $g \!\to\! \qqbar$ branchings, 
with probabilities determined up to next-to-leading logarithm
level.

In the second stage,
these partons ``hadronize", or transform into ``primary" hadrons, 
a step that is not understood quantitatively.
The ansatz of local parton-hadron duality (LPHD)~\cite{mlla}, 
that inclusive distributions of primary hadrons are the same up to a
scale factor as those for partons,
allows MLLA QCD to predict properties of distributions of 
the dimensionless variable $\xi \!=\! -\ln x_p$ for different hadrons.
Here, $x_p \!=\! 2\pstar/\ecm$ is the scaled momentum, and
\pstar and \ecm are the hadron momentum and the \epem energy, 
respectively, in the \epem center-of-mass (CM) frame.
Predictions include the shape of the $\xi$ distribution
and its dependence on hadron mass and \ecm.
At sufficiently high $x_p$, 
perturbative QCD has also been used to calculate the \ecm dependence
of the $x_p$ distributions~\cite{kkp}.

In the third stage,
unstable primary hadrons decay into more stable particles,
which can reach detector elements.
Although proper lifetimes and decay branching fractions have been
measured for many hadron species, 
these decays complicate fundamental measurements because many of the
stable particles are decay products rather than primary hadrons.
Previous measurements at \epem colliders~\cite{bohrer}
indicate that decays of vector mesons, strange baryons, and decuplet baryons
produce roughly two thirds of the stable particles;
scalar and tensor mesons and radially excited baryons have also been
observed and contribute additional secondary hadrons.
Ideally one would measure every hadron species and distinguish primary
hadrons from decay products on a statistical basis.
A body of knowledge could be assembled by reconstructing increasingly 
heavy states
and subtracting their known decay products from the measured rates of 
lighter hadrons.
The measurement of the stable charged hadrons constitutes a first step
in such a program.

There are several phenomenological models of hadronic jet production.
To model the parton production stage,
the HERWIG~5.8~\cite{herwig}, JETSET~7.4~\cite{jetset} and 
UCLA~4.1~\cite{ucla} event generators rely on combinations of
first-order matrix elements and parton-shower simulations.
For the hadronization stage, 
the HERWIG model splits the gluons produced in the first stage into 
\qqbar pairs,
combines these quarks and antiquarks locally to form colorless
``clusters", 
and decays the clusters into primary hadrons.
The JETSET model represents the color field between the partons by a 
``string'', 
and breaks the string according to an iterative algorithm into several 
pieces, each corresponding to a primary hadron.
The UCLA model generates whole events according to weights derived
from phase space 
and Clebsch-Gordan coefficients.
Each model contains free parameters controlling various aspects
of the hadronization process, whose values have been tuned to
reproduce data from \epem annihilations.
With a large number of parameters, 
JETSET has the potential to model many hadron species in detail,
whereas UCLA and HERWIG seek a more global description with fewer
parameters, including only one or two that control the relative rates
of different species.

The scaling properties, or \ecm dependences, 
of hadron production are of particular interest.
Since the process is governed by QCD, 
it is expected to be scale invariant, 
i.e. distributions of $x_p$ should be independent of \ecm
except for the effects of hadron masses/phase space and the running of \as.
The quark flavor composition varies with \ecm, 
and may also have substantial effects.
Mass effects are observed to be large unless $x_p \!\gg\! m_h/\ecm$, 
where $m_h$ is the mass of the hadron in question, 
although current experimental precision is limited at lower energies.
At high $x_p$,
the expected scaling violations have been calculated~\cite{kkp}
and found to be consistent with available data,
but experimental precision is limited for specific hadron species.
The scaling violation for inclusive charged tracks has been used to
extract \as under a number of assumptions about the dependence on
event flavor and particle type~\cite{delphias}.
Improved precision at 10.54~\gev
would provide stringent tests of such assumptions and more
robust measurements of \as.

The production of the charged hadrons \pipm, \Kpm, and \ppbar has been
studied in \epem annihilations at \ecm values of 
10~\gev~\cite{argus89},
29~\gev~\cite{pikptpc},
34 and 44~\gev~\cite{pikptasso},
58~\gev~\cite{pikptopaz},
91~\gev~\cite{pikpdelphi,pikpopal,pikpaleph,pikpsld},
and at several points in the range 130--200~\gev~\cite{pikpdelpiw}.
Recently, Belle has measured \pipm and \Kpm production at
10.52~\gev~\cite{belle13}.
Results for 91~\gev, near the $Z^0$ pole,
include precise measurements in inclusive hadronic events, 
as well as measurements for separated quark flavors,
quark and gluon jets,
and leading particles~\cite{lpsld,lpopal}.
The higher- and lower-energy measurements are, however, 
limited in precision and $x_p$ coverage.
Improved precision over the full $x_p$ range at 10.54~\gev would probe
the large scaling violations in detail and provide sensitive new tests
of QCD calculations and hadronization models.

In this article, 
we present measurements of the inclusive normalized production cross
sections of charged pions, kaons, and protons per \eeqqb event.
We use 0.91~\invfb of data recorded by the \babar\ detector at the
PEP-II storage ring at SLAC in March, 2002, at a CM energy of 10.54~\gev.
This is a small fraction of the \babar\ ``off-resonance'' data, 
recorded during a period dedicated to the delivery of stable beams
and constant luminosity.
The detector experienced relatively low backgrounds and ran in its
most efficient configuration, 
which was not changed in this period.
In parallel, we analyze 3.6~\invfb of data recorded at the \Y4S
resonance (10.58~\gev) during the remainder of this period,
February--April, 2002.
This ``on-resonance'' sample provides independent, stringent
systematic checks,
and the combined samples provide data-derived calibrations of the
tracking and particle identification performance.
The uncertainties on the results are dominated by systematic
contributions.

The detector and event selection are described in
sections~\ref{sec:detector}--\ref{sec:selection}.
The selection of high quality charged tracks and their identification as
pions, kaons or protons is discussed in section~\ref{sec:pid}.
The measurement of the cross sections, 
including corrections for the effects of 
backgrounds, detector efficiency and resolution, and the boost of the
\epem system in the \babar\ laboratory frame, 
are described in section~\ref{sec:frax}.
The results are compared with previous results and with the
predictions of QCD and hadronization models in
section~\ref{sec:results},
and are summarized in section~\ref{sec:summary}.

\section{The \babar\ Detector}
\label{sec:detector}

The \epem system is boosted in the \babar\ laboratory frame by 
$\beta\gamma = 0.56$ along the $e^-$ beam direction.
We call this direction ``forward'', $+z$,
and denote quantities in the \epem CM frame with an asterisk, 
and those in the laboratory frame with a subscript `lab'.
For example, 
\pstar denotes the magnitude of a particle's momentum in the CM frame
and \thstr its angle with respect to the $e^-$ beam direction, 
and \plab and \thlab denote the corresponding quantities in the
laboratory frame.
For \eeqqb events at $\ecm \!=\! 10.54$~\gev, 
the maximum \pstar value is $\ecm/2=5.27$~\gevc,
but the maximum \plab value depends on polar angle, 
with values of 3.8~\gevc at $\cthlab \!=\! -0.8$ and 
7~\gevc at $\cthlab \!=\! +$0.9. 
Thus, particles with a given \pstar value have different \plab values
in different regions of the detector, 
and are measured with different efficiencies and systematic
uncertainties.

The \babar\ detector is described in detail in Ref.~\cite{babarNIM}. 
In this analysis, 
we use charged tracks measured in the silicon vertex tracker (SVT) 
and the drift chamber (DCH),
and identified in the DCH and the detector of internally reflected
Cherenkov light (DIRC).
We also use energy deposits measured in the CsI(Tl) crystal
calorimeter (EMC) to identify electron tracks and construct quantities
used in the event selection.
These subdetectors operate in a 1.5~T solenoidal magnetic field.

The SVT comprises five double-sided layers of strip detectors,
each of which measures a coordinate along ($z$) and azimuthally around
($\phi$) the beam axis.
The DCH includes 40 layers of axial and stereo wires.
Their combined resolution is
$\sigma_{p_t} / p_t = 0.45\% \oplus (0.13\% \cdot p_t[\gevc])$,
where $p_t$ is the momentum transverse to the beam axis.
The DCH measures ionization energy loss (\dEdx) with a resolution of
8\%.

The DIRC~\cite{dircNIM} consists of 144 fused silica radiator bars
that guide Cherenkov photons to an expansion volume filled with water
and equipped with 10,752 photomultiplier tubes.
It covers the polar angle range $-0.8 \!<\! \cthlab \!<\! 0.9$.
The refractive index of 1.473 corresponds to Cherenkov thresholds of 
0.13, 0.48 and 0.87~\gevc for \pipm, \Kpm and \ppbar, respectively.
The Cherenkov angles of detected photons are measured with an
average resolution of 10.2~\mrad.
Tracks with very high \plab 
yield an average of 20 detected photons at $\cthlab \!=\! 0$, 
rising to 65 photons at the most forward and backward angles.

The EMC comprises 5,760 CsI(Tl) crystals in a projective geometry
that measure clusters of energy with a resolution of 
$\sigma_{E} / E = 1.85\% \oplus (2.32\%/^4\sqrt{E[\gev]})$,
An algorithm identifies electrons using track momentum combined with
EMC measurements of energy and shower shape.
It has better than 95\% efficiency for $\plab \!>\! 0.2~\gevc$,
and hadron misidentification rates of up to 1\% for
$\plab \!<\! 0.5~\gevc$ and at most 0.1\% for higher momenta.

\section{Hadronic Event Selection}
\label{sec:selection}

The event selection is optimized for low bias across
the hadron momentum spectra and \eeqqb event multiplicity distribution,
while minimizing backgrounds from other physics processes
and beam-wall and beam-gas interactions.
After fitting each combination of three or more reconstructed charged
tracks to a common vertex,
we require: 
\begin{enumerate}
\item
 at least three charged tracks and one good vertex, where a good
 vertex has a $\chi^2$ confidence level above 0.01;
\item
 the good vertex with the highest track multiplicity to lie
 within 5~\mm of the beam axis, and within 5~\cm of the center of the 
 collision region in $z$;
\item
 the second Fox-Wolfram moment~\cite{foxwolfram} to be less than 0.9;
\item
 the sum of the energies of the charged tracks and unassociated
 neutral clusters \etot to be in the range 5--14~\gev;
\item
 the polar angle of the event thrust~\cite{thrust} axis in the CM
 frame to satisfy $|\cththr | <0.8$; 
\item
 the track with the highest \plab not to be identified as an electron
 in events with fewer than six tracks,
 and neither of the two highest-\plab tracks to be identified as an
 electron in events with only three tracks.
\end{enumerate}
Criteria 3 and 6 reject leptonic events, 
$\epem \!\to\! \epem$, $\mup\mu^-$, and \taup\taum.
Criteria 4 and 5 ensure that the event is well contained within
the sensitive volume of the detector, 
resulting in smaller corrections and lower biases.
These criteria select 2.2 million events in our off-resonance signal
sample
and 11.8 million events in our on-resonance calibration sample.
About 27\% of the events in the latter sample are \Y4S decays.

\begin{figure}[tbp]
 \begin{center}
  \includegraphics[width=0.87\hsize]{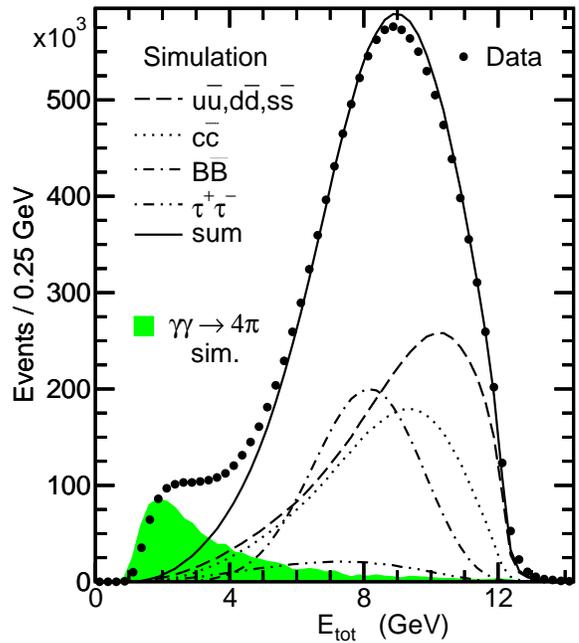}
  \caption{
   Distributions of the total visible energy per event,
   after all other selection criteria have been applied, 
   in the on-resonance data and simulation.
   The sum of the hadronic and $\tau$-pair simulations is normalized
   to the data in the region above 5~\gev,
   and the $\gamma\gamma$ simulation is normalized arbitrarily. 
  }
  \label{fig:etot}
 \end{center}
\end{figure}

We evaluate the performance of the event selection using the data and
a number of simulations, 
each consisting of a generator for a certain type of event combined
with a detailed simulation of the \babar\ detector using the 
GEANT4~\cite{geant4} package.
For signal \eeqqb events, 
we use the JETSET~\cite{jetset} event generator and obtain simulated
selection efficiencies of
0.68 for $u\bar{u}$, $d\bar{d}$ and $s\bar{s}$ events, and
0.73 for $c\bar{c}$ events.
As cross checks, 
we also use the UCLA model combined with GEANT4,
and the JETSET, UCLA and HERWIG models with a fast detector simulation
and several different parameter sets.
These give efficiency variations of at most 0.5\%.
In all cases, the largest signal loss is due to the requirement on
\ththr, 
which ensures that the event is well contained within the
sensitive volume of the detector, 
resulting in low \pstar and multiplicity biases.
We find consistency between data and simulation in a number of
distributions of event and track quantities;
the largest discrepancy we observe is a possible shift in the \etot
distribution (see Fig.~\ref{fig:etot}),
which could indicate an efficiency difference of at most 0.5\%.

We use the KORALB~\cite{koralb} generator to simulate $\mu$- and
$\tau$-pair events.
The former provide a negligible contribution, 
but the latter are the largest source of background, 
estimated to be 4.5\% of the selected events 
and to contribute up to 25\% of the charged tracks at the highest
momenta. 
However, 
the relevant properties of $\tau$-pair events are well
measured~\cite{pdg},
and their contributions can be simulated and subtracted reliably.

Radiative Bhabha events ($\epem \!\to\! \epem\gamma$) are an
especially problematic background, 
as their cross section in the very forward and backward regions
is larger than the \qqbar cross section and varies rapidly with
\cthstr.
Bremsstrahlung, photon conversions, and other interactions in the
detector material are difficult to simulate in these regions, 
and can result in events with 3--6 tracks, 
most of which are from electrons or positrons.
Simulations using the BHWIDE~\cite{bhwide} generator predict that
these events are reduced to a negligible level by criteria 1--5
plus a requirement that the highest-\plab track in the 3- and 4-track 
events not be identified as an electron.
However, a comparison of \ep and \en angular distributions in the
selected data indicates a larger contribution.
Therefore, we impose the tighter \epm vetoes given in criterion 6,
and estimate from the data a residual radiative Bhabha event
contribution of 0.1\% of the selected events and 
up to 8\% of the charged tracks at our highest momenta and $|\cthlab|$
values. 

Initial-state radiation (ISR), 
$\epem \!\to\! \gamma\epem \!\to\! \gamma\qqbar$, produces hadronic
events with a lower effective CM energy.
Low-energy ISR photons are present in all events and are simulated
adequately in the JETSET model.
The event selection is designed to suppress events with higher-energy
ISR photons, including
radiative return to the \Y1S, \Y2S and \Y3S resonances
(whose decays have very different inclusive properties from \eeqqb
events) 
and events with a very energetic ISR photon recoiling against a hadronic
system, which can mimic 2-jet events.
Using the AFKQED generator~\cite{afqed}, 
we find that the combination of the requirements on \etot and
\ththr reduces the energetic-ISR background to negligible levels, 
and the $\Upsilon(nS)$ background to one event in $10^5$.

We use the GAMGAM~\cite{gamgam} generator to study backgrounds from
2-photon ($\gamma\gamma$) processes,
$\epem \!\to\! \epem\gamma\gamma \!\to\! \epem+$hadrons.
Neither the total cross section nor those for any specific final
states are known,
but such events have relatively low track multiplicity and \etot
since the final-state \epm and some of the hadrons generally go
undetected along the beam direction. 
The \etot distribution for events in the data satisfying all other
selection criteria is shown in Fig.~\ref{fig:etot}.
It features a structure in the 1-5~\gev range that is not described
by the signal plus $\tau$-pair simulations,
but can be described qualitatively by the addition of $\gamma\gamma$
events.
Since the mixture of final states is unknown, 
we consider $\gamma\gamma \!\to\! \pip\pim\pip\pim$,
which has the largest fraction of events with \etot$>$5~\gev of any
final state with at least three tracks.
The simulated \etot distribution is shown as the shaded histogram in
Fig.~\ref{fig:etot}.
If normalized to account for the entire excess in the data,
such events would make up less than 1\% of the selected sample
(5$<$\etot$<$14~\gev),
with a track momentum distribution similar to that in $\tau$-pair events.
We take this as an upper limit on our $\gamma\gamma$ background
and vary its contribution over a wide range in evaluating the 
systematic uncertainty,
as discussed in Sec.~\ref{sec:bkg}.

Backgrounds from beam-gas and beam-wall interactions can be studied
using distributions of event vertex position in the data.
From the distribution in distance from the beam axis for events
satisfying all selection criteria except those on the vertex position, 
we conclude that the beam-wall background is negligible.
From the distribution in $z$ after including the requirement that the
vertex be within 5~mm of the beam axis,
we estimate that four beam-gas events are selected per $10^5$ signal
events.
We neglect both of these backgrounds.

We consider a number of other possible backgrounds, 
including two-photon events with one or both \epm detected
and 
other higher-order quantum electrodynamics (QED) processes producing
four charged leptons or two leptons and a \qqbar pair;  
all are found to be negligible.
We estimate that the selected sample is 95.4$\pm$1.1\% pure in
\eeqqb events,
with the background dominated by $\tau$-pairs and the uncertainty by 
$\gamma\gamma$ events.
The on-resonance calibration sample contains the same mixture of
\eeqqb and background events, plus a 27\% contribution from \Y4S
decays.

\section{Charged Track Selection and Identification}
\label{sec:pid}

The identification of charged tracks as pions, kaons or protons is
performed using an algorithm that combines the momentum and ionization
energy loss measured in the DCH and the velocity measured via the
Cherenkov angle in the DIRC.
To ensure reliable measurements of these quantities, 
we require tracks to have:
i) at least 20 measured coordinates in the DCH;
ii) at least 5 coordinates in the SVT, including at least 3 in $z$;
iii) a distance of closest approach to the beam axis of less than 1~\mm;
iv) a transverse momentum $p_t \!>\! 0.2~\gevc$;
v) a polar angle \thlab satisfying $-0.78 \!<\! \cthlab \!<\! 0.88$;
and 
vi) an extrapolated trajectory that intersects a DIRC bar.
The first criterion ensures good \dEdx resolution, 
the first three criteria select tracks from particles that originate
from the primary interaction and do not decay in flight or interact
before reaching the DIRC,
and the combination of all six criteria yields tracks
well within the DIRC fiducial volume, with
good momentum and polar angle resolution.

These criteria suppress tracks from decays of long-lived particles such as
\KS and $\Lambda$ hadrons,
which are included in many previous measurements.
Here, we report cross sections for two classes of tracks, 
denoted ``prompt'' and ``conventional".
We first measure prompt hadrons, 
defined as primary hadrons or products of a decay chain in which all
particles have lifetimes shorter than $10^{-11}$~s.
This includes products of all charmed hadron decays,
as well as those of strongly or electromagnetically decaying strange
particles,
but not those of weakly decaying strange particles.
We then obtain the conventional quantities by adding the decay
daughters of particles with lifetimes in the range 
1--3$\times$10$^{-11}$~s, 
i.e., \KS and weakly decaying strange baryons.
For this we use existing measurements of \KS and strange
baryon production~\cite{cleohad,argusk0lam}.
Either or both cross sections can be compared with other measurements,
and used to test QCD and model predictions.

In selected simulated events,
these criteria accept 82\% of the prompt charged particles generated
within the target \thlab range and with $p_t \!>\! 0.2~\gevc$.
This efficiency rises slowly from 80\% at $\plab \!=\! 0.2$~\gevc to
86\% at the highest momentum,
and is almost independent of particle type, polar angle, event flavor,
and track multiplicity.
Corrections to the simulation are discussed in Sec.~\ref{sec:effic}.

Since the \epem system is boosted in the laboratory frame,
we divide the selected tracks into six regions of $\cthlab$: 
[$-$0.78,$-$0.33], [$-$0.33,0.05], [0.05,0.36], 
[0.36,0.6], [0.6,0.77] and [0.77,0.88], 
denoted $\theta 1$ to $\theta 6$, and analyze each region separately.
These correspond to regions of roughly equal width in \cthstr
between $-$0.92 and $+$0.69.
The tracks in each region arise from the same underlying \pstar
distribution, 
but are boosted into different ranges of \plab.
Also, heavier particles are boosted to higher \cthlab,
with low-\pstar protons and kaons populating the forward \cthlab
regions preferentially.
Thus we perform multiple (up to six) measurements for each \pstar
value,
each from a different \plab range and in a different region of the
detector. 
Their comparison provides a powerful set of cross checks on 
detector performance and material interactions,
backgrounds, 
the true \thstr and \pstar distributions,
and the boost value itself.

\subsection{Charged Hadron Identification}
\label{sec:pida}

The \dEdx measurement from the DCH provides very good separation between
low-\plab particles, 
i.e., between \Kpm and \pipm (\ppbar and \Kpm) below about 0.5
(0.8)~\gevc.
There is also modest separation, of 1--3 standard deviations ($\sigma$), 
in the relativistic rise region above about 2~\gevc,
and the separation varies rapidly at intermediate \plab.
For each accepted track, we calculate a set of five likelihoods 
$L_i^{\rm DCH}$, $i=e, \mu, \pi, K, p$,
each reflecting the degree of consistency of its measured \dEdx value
with hypothesis $i$.

The Cherenkov angle measurement from the DIRC provides very good
separation between particles with \plab between the Cherenkov
threshold and the resolution limit of
about 4~\gevc for \pipm vs.\ \Kpm and 6.5~\gevc for \Kpm vs.\ \ppbar.
The number of expected photons varies rapidly with \plab just above
threshold,
and the number detected for each track provides additional
information.
A track can be classified as being below threshold by counting the
detected photons at the angles expected for each above-threshold
particle type and comparing with the hypothesis that only background
is present. 
To make full use of this information, 
we maximize a global likelihood for the set of reconstructed tracks in
each event, 
which considers backgrounds, 
photons that could have been emitted by more than one track,
and multiple angles from a given track. 
For each track, 
we calculate a set of five likelihoods $L_i^{\rm DIRC}$, 
$i=e, \mu, \pi, K, p$,
assuming the best hypothesis for all other tracks.
These provide \Kpm-\pipm (\ppbar-\Kpm) separation that rises rapidly
with \plab from zero at the \pipm (\Kpm) Cherenkov threshold of 
0.13 (0.48)~\gevc, 
to a roughly constant value,
from which it falls off above about 2.5 (4.5)~\gevc.

\begin{figure}[tbp]
 \begin{center}
  \includegraphics[width=8.5cm]{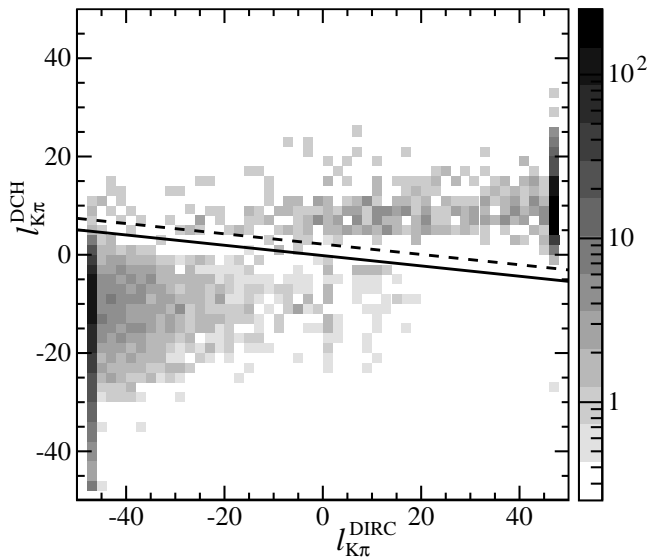}
  \caption{Simulated distribution of the $K$-$\pi$ log-likelihood
    difference $l_{K\pi}$ from the DCH vs.\ that from the DIRC for
    \pipm and \Kpm in hadronic events generated with 
    $0.6 \!<\! \plab \!<\! 0.625$~\gevc and $\cthlab \!>\! 0.05$.
    The \pipm and \Kpm are concentrated in the lower left and upper
    right regions, respectively.
    The edge bins include overflows.
    The solid (dashed) line represents an upper (lower) bound on
    identified \pipm (\Kpm).
  }
  \label{fig:dllik}
 \end{center}
\end{figure}

To make use of both DCH and DIRC information,
we consider the log-likelihood differences 
$l_{ij}^{\mbox{det}} = \ln(L_i^{\mbox{det}}) - \ln(L_j^{\mbox{det}})$, 
where det $=$ DCH, DIRC,
and we identify tracks by their positions in the 
$l_{ij}^{\rm{DCH}}$ vs.\ $l_{ij}^{\rm{DIRC}}$ planes.
The procedure is illustrated in Fig.~\ref{fig:dllik} for simulated
\pipm (lower left) and \Kpm (upper right) 
with $0.6 \!<\! \plab \!<\! 0.625$~\gevc and $\cthlab \!>\! 0.05$.
Here the DIRC provides clear separation for all but a few percent of
the tracks 
(most of the entries at the left and right edges are overflows),
but long tails are visible in the $l_{K\pi}^{\rm DIRC}$ distributions
for both \pipm and \Kpm.
The DCH separation is smaller, but the tails are shorter.
To be identified as a \pipm, a track must lie below a line in the
$l_{K\pi}^{\rm DCH}$--$l_{K\pi}^{\rm DIRC}$ plane (see
Fig.~\ref{fig:dllik}) and below another line in the
$l_{p\pi}^{\rm DCH}$--$l_{p\pi}^{\rm DIRC}$ plane.
Similarly,
an identified \Kpm lies above a line 
(dashed in Fig.~\ref{fig:dllik}) in the
$l_{K\pi}^{\rm DCH}$--$l_{K\pi}^{\rm DIRC}$ plane and below a line in the
$l_{pK}^{\rm DCH}$--$l_{pK}^{\rm DIRC}$ plane,
and an identified \ppbar lies above lines in the
$l_{p\pi}^{\rm DCH}$--$l_{p\pi}^{\rm DIRC}$ and
$l_{pK}^{\rm DCH}$--$l_{pK}^{\rm DIRC}$ planes.

The parameters describing the lines vary smoothly with \plab and \thlab,
and are optimized~\cite{brandon} to keep the misidentification rates
as low as reasonably possible, 
while maintaining high identification efficiencies that vary slowly
with both \plab and \cthlab.
The slopes are zero (i.e.\ only \dEdx information is used) for \plab 
below the lower of the two Cherenkov thresholds,
begin to decrease slowly at that threshold,
and become large and negative above about 2.5~\gevc;
although \dEdx provides some separation in this region,
the systematic uncertainties are minimized by using it only to reject
outlying tracks.
In some cases the two lines in a given plane are the same;
in most cases they are nearly parallel and separated by a few units, 
and tracks in between are not identified as any hadron type.
Fewer than 0.1\% of the tracks are identified as more than one type,
and these are reclassified as unidentified.

Electrons and muons represent only a small fraction of the tracks 
in hadronic events at $\ecm \approx 10$~\gev (at most 2\%),
and their production is understood at the level of 10\% or better
(see Sec.~\ref{sec:xssubtr}). 
They can be suppressed at this point using 
calorimeter and muon system information,
and we have done this as a cross check, obtaining consistent results.
However,
this also rejects some signal tracks, 
and the total systematic uncertainties are minimized by
including \epm and \mupm in the pion category at this stage,
and subtracting them later.
We therefore define a \emppm sample.
High-momentum \epm and almost all \mupm are indistinguishable from
\pipm in the DCH or DIRC,
so are included by the criteria noted so far.
The DIRC does separate \mupm from \pipm in a narrow \plab range near
0.2~\gevc, 
but we use only \dEdx information in this range.
To accommodate low-momentum \epm, 
we include tracks with \plab below 2~\gevc that satisfy requirements
in the 
$l_{e \pi}^{\rm DCH}$--$l_{e \pi}^{\rm DIRC}$ and
$l_{e K}^{\rm DCH}$--$l_{e K}^{\rm DIRC}$ planes.

\begin{figure*}[p]
 \begin{center}
  \includegraphics[width=17.5cm]{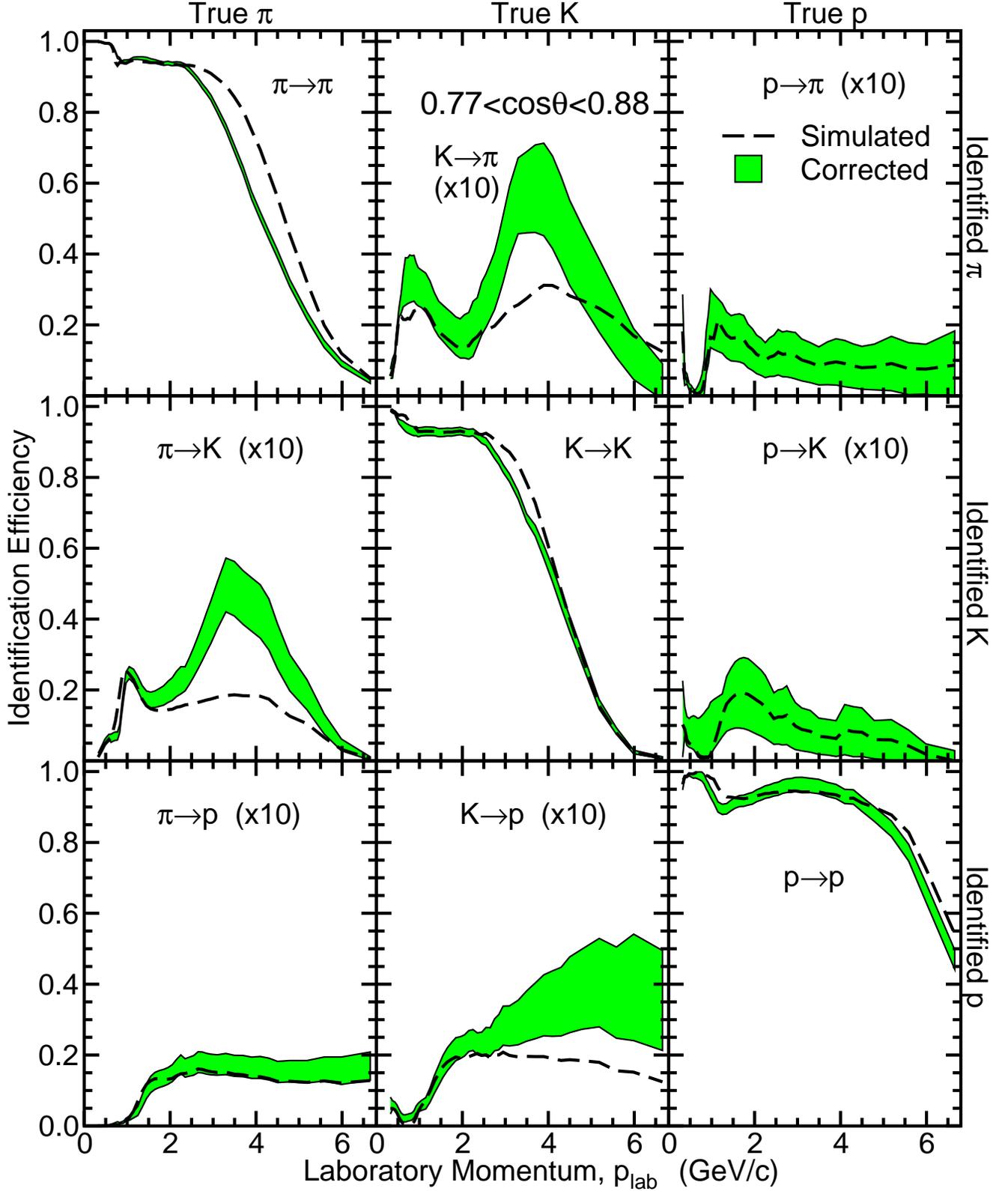}
  \caption{The simulated (dashed lines) and corrected (gray/green bands) 
           efficiency matrix for the most forward polar angle region,
           $\theta 6$, where $0.77 \!<\! \cthlab \!<\! 0.88$.
           The widths of the bands indicate the uncertainties derived
           from the control samples discussed in the text.
           The off-diagonal elements have been scaled up by a factor
           of ten for clarity.
  }
  \label{fig:effmatcorr}	
 \end{center}
\end{figure*}

We quantify the performance of our hadron identification procedure in
terms of a momentum-dependent identification efficiency matrix
{\boldmath $E$},  
where each element $E_{ij}$
represents the probability that a selected track from a true $i$-hadron 
is identified as a $j$-hadron, with $i,j=(e\mu\pi),K, p$.
The matrix predicted by the detector simulation for our most forward
polar angle region, $\theta 6$, which covers the widest \plab range,
is shown as the dashed lines in Fig.~\ref{fig:effmatcorr}.
The efficiencies for correct identification are predicted to be
very high at low \plab, where $\dEdx$ separation is good,
then transition smoothly to a plateau where the Cherenkov angle
provides good separation,
then fall off at higher \plab
where the Cherenkov angles for different particles converge.
The predicted probabilities for misidentifying a particle as a
different type are below 2.5\%.
Essentially all tracks are identified as some particle type at low \plab,
1--3\% are classified as ambiguous in the plateau regions,
and larger fractions are so classified as the efficiency falls off,
since we choose to maintain constant or falling misidentification 
rates.

Similar performance is predicted in the other \cthlab regions.
In $\theta 1$ and $\theta 2$, the two most backward regions, 
\plab does not exceed 3.5--4~\gevc,
so no fall off is visible in $E_{pp}$ at high \plab,
and $E_{\pi\pi}$ and $E_{KK}$ drop only to 30--70\% of their plateau
values.
Thus we are able to measure the high \pstar range well in multiple
\cthlab regions.
In the next few subsections, however, we focus on $\theta 6$, 
since it spans the widest range in efficiencies and requires the
largest corrections to the simulation.

\subsection{Calibration of the Identification Efficiencies}
\label{sec:pidb}

We calibrate the efficiency matrix from the combined off- and
on-resonance data set, 
using samples of tracks with known hadron content and characteristics
as similar as possible to our selected tracks.
For example, 
we construct $\KS \!\to\! \pip\pim$ candidates from tracks satisfying
criteria (i) and (iv)--(vi) presented at the beginning of
Sec.~\ref{sec:pid},
with a less restrictive requirement of three coordinates in the SVT
and an additional requirement that there be a coordinate from one of
the two outer layers of the DCH.
Pairs of oppositely charged tracks must have a fitted vertex more than
0.5~\cm from the beam axis,
a reconstructed total momentum direction within 50~mrad of the line
between their fitted vertex and the event vertex,
and an invariant mass in the range 486--506~\mevcc.
The percent-level non-\KS contribution is predominantly from pions,
so these tracks constitute a clean sample of \pipm that are produced
in hadronic events and cross most of the tracking system. 
In simulated events, this sample has $E_{\pi j}$ values
within 0.5\% of those of the prompt \pipm in the same events.
We calculate efficiencies from this \KS sample in both data and
simulation, 
and use their differences to correct the prompt \pipm simulation.
This sample covers \plab up to about 1.5~\gevc with high 
precision.

A similar selection of $\Lambda \!\to\! p\pim$ and 
$\overline{\Lambda} \!\to\! \pbar\pip$ 
candidates provides a sample of 0.4--3.5~\gevc \ppbar 
(and another sample of soft pions) in hadronic events.
We also reconstruct two samples of $\phi \!\to\! \Kp\Km$
decays in which either the \Kp or \Km is identified,
providing 0.2--2~\gevc \Km and \Kp samples that are subsamples of our
main sample.
These samples contain substantial backgrounds, 
and we extract $E_{pj}$, $E_{\pbar j}$, $E_{\Kp j}$ and $E_{\Km j}$
from sets of simultaneous fits to the four $\ppbar \pi^\mp$
or $\Kp\Km$ invariant mass distributions in which the \ppbar or the
other kaon is identified as a pion, kaon, proton or no type.

We obtain samples of 0.6--5~\gevc \pipm and \Kpm by reconstructing
candidate 
$D^{\star +} \!\!\to\! D^0\pip \!\!\to\! \Km\pip\pip$ (and charge conjugate)
decays and selecting those with a $\Km\pip\pip$--$\Km\pip$ mass
difference in the range 143--148~\mevcc.
The $\Km\pip$ invariant distribution shows a $D^0$ signal with a peak
signal-to-background of eleven.
These tracks are predominantly from \Y4S decays and \ccbar events, 
but have simulated $E_{Kj}$ and $E_{\pi j}$ values within 1\% and 0.5\%,
respectively, of those from all prompt \Kpm and \pipm in hadronic
events.
Requiring the \pim (\Kp) candidate track to be so identified and the
\Km (\pip) track to satisfy our selection criteria,
we evaluate $E_{\Km j}$ ($E_{\pip j}$) as the fraction of the
sideband-subtracted entries in the $D^0$ peak in which the \Km (\pip)
is identified as type $j$.

We select $\epem \!\to\! \tau^+\tau^-$ events in which one of the
$\tau$ decays contains a single charged track (1-prong) and the
other contains one or three (3-prong) charged tracks.
These tracks constitute \emppm samples that are not from a hadronic
jet environment and have different \epm:\mupm:\pipm content, 
as well as a small but well known \Kpm component.
However, these samples have simulated identification efficiencies 
within a few percent of those for \pipm in hadronic events,
and they allow us to study high-\plab tracks and tracks that
are isolated (1-prong) or relatively close together (3-prong) in the
detector.
We also apply independent electron and muon selectors to the 1-prong
sample,
in order to check that the small differences in performance between  
\epm, \mupm and \pipm are simulated correctly.

Results from the different calibration samples are consistent where
they overlap, 
as are those from positively and negatively charged tracks 
and from on- and off-resonance data.
Considering the set of constraints provided by these samples,
we derive corrections to the simulated $E_{ij}$ elements that vary
smoothly with \plab and \cthlab.
The correction to each $E_{ij}$ in each \cthlab region is a
continuous, piecewise-linear function of \plab,
with an uncertainty given by the statistically most precise
calibration sample at each point.
The resulting calibrated efficiencies in the $\theta 6$
region are shown as the gray/green bands in Fig.~\ref{fig:effmatcorr};
their centers represent the calibrated efficiencies,
and their half widths the uncertainties.

The pion efficiencies $E_{\pi j}$ 
(left  column of Fig.~\ref{fig:effmatcorr}) 
are measured well over the full \plab range, 
with corrections and uncertainties near or below the percent level 
for $\plab \!<\! 2.5$~\gevc.
There are substantial corrections to $E_{\pi\pi}$ and $E_{\pi K}$ in
the 3--5~\gevc range,
which is sensitive to the details of the DIRC geometry and
backgrounds.

The kaon efficiencies $E_{Kj}$ 
(middle column of Fig.~\ref{fig:effmatcorr}) 
are measured for $\plab \!>\! 0.4~\gevc$
with somewhat larger uncertainties than for $E_{\pi j}$.
The corrections to $E_{KK}$ and $E_{K\pi}$ are similar at most \plab
to those on $E_{\pi\pi}$ and $E_{\pi K}$, respectively, 
as expected from the near symmetry in the \dedx and Cherenkov angle
measurements.
They have opposite sign, as expected, in the region just
above kaon threshold, 0.5--1~\gevc.
The large correction to $E_{Kp}$ near 6~\gevc is consistent with the
corrections to $E_{\pi K}$ and $E_{K\pi}$
with \plab scaled by a factor of roughly 1.9, 
the ratio of the proton and kaon masses, as expected.

Below 0.4~\gevc, 
the kaon calibration samples have high backgrounds and do not yield
useful results.
However, 
the identification efficiencies are very high,
we expect strong correlations between hadron types up to
0.6~\gevc, 
and the calibration data are consistent with full correlation between
0.4 and 0.6~\gevc.
Therefore, 
we apply the same small corrections to $E_{KK}$ as for $E_{\pi\pi}$,
and to $E_{K\pi}$ and $E_{Kp}$ as for $E_{\pi K}$ at 0.2~\gevc,
with the uncertainty doubled arbitrarily to account for any incomplete
correlation.
We apply the corrections and uncertainties from the kaon calibration
samples to $E_{KK}$ and $E_{K\pi}$ at 0.6~\gevc, 
and vary the corrections and uncertainties linearly between 0.2 and
0.6~\gevc. 
Due to the higher proton mass, 
the corresponding region in $E_{Kp}$ extends to 1.0~\gevc,
so we match the corrections at that value.

The proton efficiencies $E_{pj}$ 
(right column of Fig.~\ref{fig:effmatcorr}) 
are measured well in the range 
0.8--3.5~\gevc, 
and the corrections show the expected correlations with the other
elements.  
Again, we expect complete correlations at low \plab,
and we apply the same corrections to $E_{p\pi}$, $E_{pK}$, and $E_{pp}$ 
as for $E_{\pi p}$, $E_{Kp}$, and $E_{KK}$, respectively, at 0.2~\gevc,
with doubled uncertainties.
We then match them to their respective proton calibration values at
1~\gevc.
Above 3.5~\gevc, the statistical precision of the proton calibration
sample is limited, and we exploit the correlation expected between 
$E_{pp}$ in the 2--6.5~\gevc range, 
and $E_{\pi\pi}$ and $E_{KK}$ in the corresponding 1.1--3.4~\gevc range.
The three corrections are consistent in the lower part of this range,
and in the upper part
we average the corrections to $E_{\pi\pi}$ and $E_{KK}$, 
scale them up in \plab,
and apply them to $E_{pp}$ with an uncertainty twice that on the
$E_{KK}$ correction.
We match to the proton calibration sample at 3.1~\gevc, 
where the uncertainties from the two approaches are comparable. 

Due to the low value of the proton fraction, 
the criteria for proton identification are more stringent than for 
pion or kaon identification at high \plab, 
so that $E_{p\pi}$ and $E_{pK}$ are smaller than the other
misidentification rates, as are the corrections.

Corrections to the efficiencies in the other \cthlab regions are
similar in form and generally smaller than those shown in
Fig.~\ref{fig:effmatcorr}.
Even though the uncertainties of some misidentification rates 
are relatively large,
they result in small systematic uncertainties of the result,
since the rates themselves are sufficiently low.
The uncertainties of the correct identification efficiencies are
important, especially at high \plab.
However, 
high-\pstar particles are measured well in the more backward
\cthlab regions,
and the final result is an average over the six regions.

\section{Measurement of the Differential Cross Sections}
\label{sec:frax}

The objects of this measurement are the production cross sections 
per unit \pstar, $(1/\sigma^{\rm had}_{\rm tot}) \, (d\sigma_i /d\pstar$),
$i \!\!=\! \pi, K, p$,
normalized to the total hadronic event cross section 
$\sigma^{\rm had}_{\rm tot}=3.39~\nb$ at our CM energy of 10.54~\gev.
We present these in the equivalent and conventional form
$(1/N_{\rm evt}) \, (dn_i /d\pstar)$, 
where $N_{\rm evt}$ and $n_i$ are the numbers of hadronic events and
$i$-particles, respectively.
 
From our samples of identified \pipm, \Kpm and \ppbar, 
we use the corrected identification efficiency matrices described in
the preceding section to construct the raw production rates
$(1/N_{\rm evt}^{\rm sel}) \, (dn_i / d\plab)$, $i \!\!=\! (e\mu\pi), K, p$,
defined as the numbers of reconstructed particles
per selected event
per unit momentum in the laboratory frame.
We subtract backgrounds and apply corrections to account for the
effects of detector efficiency and resolution, 
and the event selection procedure.
We do this separately in each of the six \cthlab regions,
and also in the on-resonance sample for control purposes.

We transform each corrected rate into a cross section in the \epem CM
frame, 
where we compare and combine the results from the six \cthlab 
regions.
Subtracting the expected contributions from leptons, 
we obtain our prompt results, $(1/N_{\rm evt})(dn_i^{\rm prompt}/d\pstar)$. 
We add the expected contributions from decays of \KS and weakly decaying 
strange baryons to obtain conventional cross sections,
and we calculate ratios of cross sections and charged hadron fractions.
Each of these steps is described in detail in the following subsections,
and each involves a number of systematic checks and uncertainties.
The systematic uncertainties are summarized in the final subsection.

\subsection{Cross Sections in the Laboratory Frame}
\label{sec:xslab}

\begin{figure}[tbp]
 \begin{center}
  \includegraphics[width=\hsize]{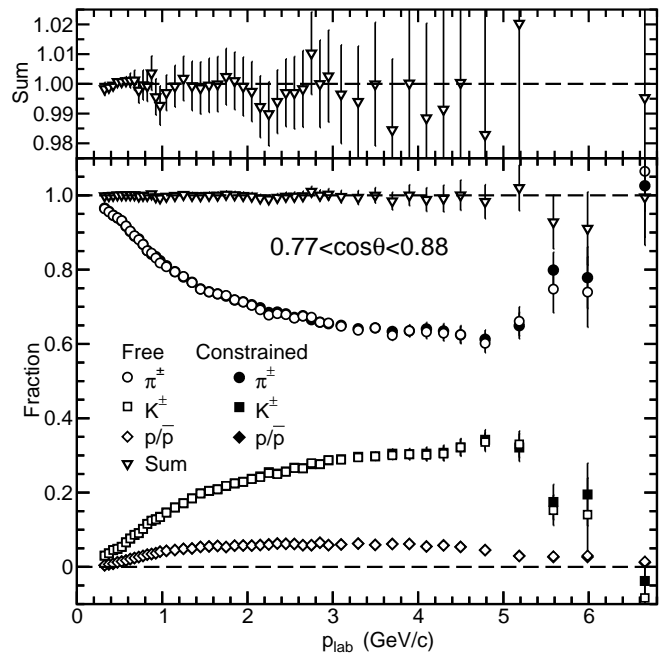}
  \caption{ 
   \label{fig:frxconst}
  Raw \pipm (circles), \Kpm (squares) and \ppbar (diamonds)
  fractions measured in $\theta 6$, the most forward \cthlab region.
  The solid (open) symbols represent the results with (without)
  the constraint that they sum to unity in each \plab bin.
  They are indistinguishable in most cases.
  The sums of unconstrained fractions are shown as the triangles,
  and in an expanded view in the upper plot.
  The error bars include statistics and the systematic uncertainties
  arising from the calibration of the particle identification
  efficiencies. 
  }
 \end{center}
\end{figure}

In each \plab bin, we count $n_j$, 
the number of tracks identified as type $j \!=\! (e\mu\pi), K, p$.  
These can be related to the true fractions $f_i$ of tracks that are of 
type $i$ by $n_j = n \Sigma_i E_{ij} f_i$, 
where $n$ is the total number of selected tracks and the efficiency
matrix {\boldmath $E$} is described in Sec.~\ref{sec:pid}.
We first solve this set of equations in each bin for the three $f_i$ values,
and check that their sum is consistent with unity.
This check is sensitive to many systematic effects on {\boldmath $E$}, 
and if we apply no corrections to the simulated {\boldmath $E$}, 
we find significant differences from unity in several places,
most notably in the DCH-DIRC crossover region near 0.7~\gevc
and at the highest momenta in the forward polar angle regions.
The on-resonance control sample shows the same differences.
After the corrections, 
the sum is consistent with unity in all bins within the systematic
uncertainties obtained by propagating the uncertainties on the nine
$E_{ij}$.
The fractions and their sum in the most forward \cthlab region,
$\theta 6$, are shown in Fig.~\ref{fig:frxconst}. 
Neighboring points are correlated due to the efficiency correction
procedure.

We then recalculate the fractions with the added constraint that their
sum be unity.
The recalculated fractions are also shown in Fig.~\ref{fig:frxconst},
and are almost indistinguishable from the unconstrained fractions.
In the systematic error propagation,
we account for the constraint by varying the three efficiencies
$E_{jj}$ independently,
and in each case varying both corresponding misidentification rates
$E_{jk}$ in the opposite direction.
Both the statistical and systematic uncertainties decrease slightly
with the addition of the constraint.
It also introduces strong statistical correlations between the three
particle types,
but since the results are dominated by systematic effects, 
we neglect these.

Several additional systematic checks are performed, 
including varying the misidentification rates by three times their
uncertainties,
changing the \plab ranges over which we fit the corrections to
{\boldmath $E$},
using different event flavor mixtures in the simulation,
and using the efficiencies measured in the control samples directly, 
rather than using them to correct the simulation. 
We find no change in the results larger than the relevant systematic 
uncertainty.

Each fraction is multiplied by the number of accepted tracks in that bin
and divided by the number of selected hadronic events and by the 
bin width to obtain raw normalized cross sections.

\subsection{Background Subtraction}
\label{sec:bkg}

\begin{figure}[tbp]
 \begin{center}
  \includegraphics[width=\hsize]{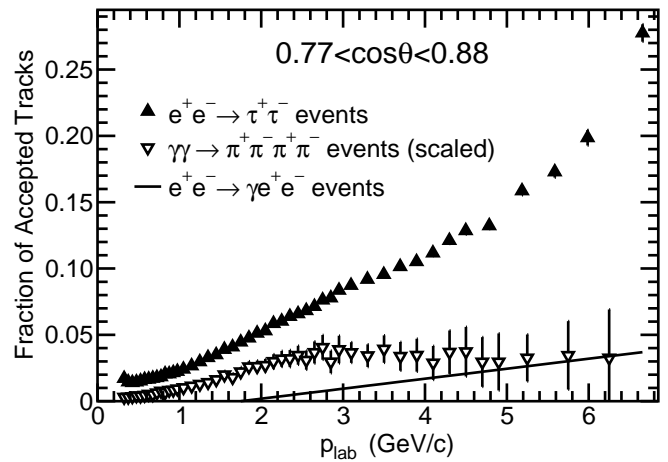}
  \vspace*{-5pt}
  \caption{ \label{fig:bkgtau}
   Fractional contributions to the selected track sample from
   $\tau$-pair (up triangles), $\gamma\gamma \!\to\! 2\pip 2\pim$ 
   (down triangles) and radiative Bhabha (line) events in $\theta 6$,
   as functions of the measured \plab.
   The $\gamma\gamma \!\to\! 2\pip 2\pim$ cross-section is scaled as
   discussed in the text, and represents an upper bound.
   }
 \end{center}
\end{figure}

We subtract backgrounds due to other physics processes, 
interactions in the detector material,
and strange-particle decay products.
As discussed in Sec.~\ref{sec:selection},
there are three physics processes with non-negligible background
contributions to our event sample: 
$\tau$-pair, two-photon and radiative Bhabha events.
Figure~\ref{fig:bkgtau} shows the simulated fractional contributions
to the selected tracks in region $\theta 6$ from these three sources.

The contribution from $\tau$-pair events is small at low \plab,
but grows steadily to over 20\% at higher momenta.
There are similar contributions in the other \cthlab regions.
The simulation of $\tau$-pair production and decay is reliable at the
sub-percent level,
and our detector simulation is reliable
(after the corrections described in Sec.~\ref{sec:effic}) to 1--2\%.
However, 
since we normalize per selected event,
we must consider the relative event selection efficiency.
Here, our simulation is also quite reliable for $\tau$-pairs,
but less so for hadronic events, discussed in Sec.~\ref{sec:systematics},
and the uncertainty corresponds to a roughly constant 10\% relative
uncertainty on the tracks from $\tau$-pair events.
We therefore subtract the absolute prediction of the simulation with a
10\% relative uncertainty.

The contribution from two-photon events is not well understood,
but we can set an upper limit by scaling our simulated
$\gamma\gamma \!\to\! 2\pip 2\pim$ sample to account for the
structure observed at low \etot, 
discussed in Sec.~\ref{sec:selection} and shown in Fig.~\ref{fig:etot}.
The resulting contribution is shown by the triangles in
Fig.~\ref{fig:bkgtau}.
Due to the kinematics of $\gamma\gamma$ events and the detector
acceptance, this background is highest in $\theta 6$, 
somewhat smaller in $\theta 1$,
and about half as large in the central regions.
Most $\gamma\gamma$ events contain more charged and neutral hadrons
than the $2\pip 2\pim$ final state, 
some of which are outside the acceptance,
yielding smaller values of \etot.
Therefore, 
we expect to select far fewer events than indicated by this sample,
containing mostly lower-\plab tracks,
and Fig.~\ref{fig:bkgtau} shows a substantial overestimate at high
\plab and an upper bound at lower \plab.
This limit is at most 4\% and well below the $\tau$-pair contribution,
so we make no correction, 
but assign a systematic uncertainty corresponding to one half of the
limit in each bin.

As discussed in Sec.~\ref{sec:selection},
the simulation predicts a negligible contribution from radiative
Bhabha events,
but may be unreliable, 
especially in the forward and backward directions.
Due to the $t$-channel contribution to their production process, 
such events would exhibit a charge asymmetry with a characteristic
dependence on \plab and \cthlab.
In our selected \emppm sample,
we observe significant differences between positively and negatively
charged tracks that reach 10\% and $-$4\% at the highest \plab
in the most backward and forward \cthlab regions, respectively,
and show an angular dependence consistent with radiative Bhabha events.
We make a smooth parametrization of this difference, and subtract it
from our $(e\mu\pi)$ cross section.
The effect is a few percent at high momenta in the forward 
(see Fig.~\ref{fig:bkgtau}) and backward \cthlab regions, 
but below 1\% in the central regions.
This procedure also accounts for any residual events from
$\epem \!\to\! \epem\epem$ or other higher-order QED processes with
forward-peaking cross sections.

After subtracting these $\tau$-pair and radiative Bhabha backgrounds, 
we normalize by the estimated number of hadronic events in the
selected sample, 
to obtain background-subtracted differential cross sections.

\begin{figure}[tbp]
 \begin{center}
  \includegraphics[width=\hsize]{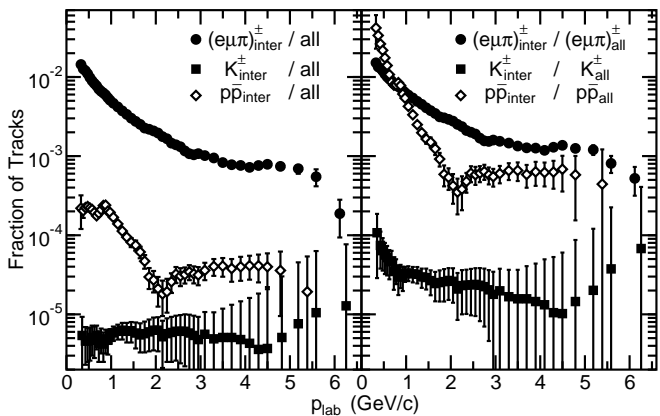}
  \caption{ 
   \label{fig:detint}
   Simulated fractional contributions to 
   the overall selected track sample (left) and 
   the selected tracks of the same type (right)
   from pions (circles), kaons (squares) and protons (diamonds)  
   produced through interactions in the detector material.
   Results for tracks in the on-resonance sample in $\theta 6$
   are shown, and the point-to-point variations have been smoothed.
  }
 \end{center}
\end{figure}

Interactions of particles with the detector material can lead to
tracking inefficiencies, which are discussed in Sec.~\ref{sec:effic}, 
and also to the production of extraneous charged tracks that satisfy
the signal-track criteria.
Most interaction products fail the selection criteria, 
but two categories require care: 
a highly asymmetric photon conversion can produce an electron or
positron that points back to the event vertex;
and a pion interacting with a nucleon through a $\Delta$ resonance can
produce a proton nearly collinear with the pion.
Figure~\ref{fig:detint} shows the simulated fractional contributions
from interaction products.
Photon conversions account for the vast majority,
as much as 1.5\% of the selected tracks at the lowest \plab value,
but well below 1\% over most of the \plab range.
We have measured the photon conversion rate in our data,
and the simulated rate lies within 20\% of this rate for all \plab,
so we subtract the simulated fractional contribution to the
$(e\mu\pi)$ sample,
shown in the right plot of Fig.~\ref{fig:detint}.
Since the measurement uses conversions with two tracks that fail our
selection criteria,
and the chance of passing depends on details of the detector
simulation,
we assign an arbitrary and conservative systematic uncertainty equal
to 50\% of the correction.

Protons produced in the detector material represent a small fraction
of all selected tracks,
but as much as 4\% and 15\% of those identified as protons in the
lowest kinematically allowed \plab bins in $\theta 6$
(shown in Fig.~\ref{fig:detint}) and $\theta 1$, respectively.
There is a concentration of material in the \babar\ detector between
the SVT and the DCH,
and protons produced in this region can be studied using tracks that are
identified by our algorithm as protons, 
but have measured \dEdx in the SVT inconsistent with a proton and
consistent with a pion.
Our study revealed a problem with the simulation of the $\Delta$
resonances in our version of GEANT, 
for which we apply a correction.
We subtract the corrected simulated contributions of such protons 
and apply a uniform 50\% relative uncertainty, 
which is slightly larger than the statistical uncertainty on the study
in each \cthlab region.

Very few antiprotons are produced in material interactions,
but they suffer from similar uncertainties in the loss rate
(see Sec.~\ref{sec:effic}).
We measure $p$ and \pbar cross sections separately, and the results
are consistent within these systematic uncertainties.
The simulation predicts a very small number of kaons from detector
interactions, 
and we subtract the predicted fraction with an arbitrary  50\%
uncertainty.
The simulation also includes tracks arising from beam-related
backgrounds and noise in the detector,
by overlaying untriggered events from beam crossings close in time to
triggered events.
These are a small fraction of the tracks in Fig.~\ref{fig:detint},
and are included in the correction.

There are also residual tracks in the sample from weak decays of
strange particles that we must exclude from our prompt sample.
We evaluate these by reweighting our simulated \KS and \KL spectra to
reproduce the average of the measured \KS spectra at or near our CM
energy~\cite{cleohad,argusk0lam},
reweighting our simulated $\Lambda$ spectrum to match the measured
spectrum~\cite{cleohad,argusk0lam},
and applying the same weights to our simulated $\Sigma$ baryon
spectra.
The weighted simulation predicts that at the lowest \plab,
about 2\% of the selected \emppm tracks are from \KS decays and a
further 3\% from strange baryon decays,
with both contributions falling rapidly as \plab increases.
About 13\% of the selected \ppbar tracks in the lower half of the
\plab range are from strange baryon decays,
and this falls slowly toward 4\% at the highest \plab.
There are also smaller contributions of \emppm from \KL, \Kpm and
\pipm decays, 
and \Kpm from $\Omega^-$ decays.

We subtract the simulated fractions of these tracks from each cross
section,
and assign systematic uncertainties to the \KS and strange baryon
contributions based on the uncertainties on the corresponding
measured spectra~\cite{cleohad,argusk0lam}. 
The assigned uncertainties are parametrized with smooth functions that
vary with \pstar between 5\% and 35\% over the bulk of distributions,
and increase toward 100\% at zero and the kinematic limits,
where the contributions vanish.
We assign an arbitrary 50\% relative uncertainty to all other sources.

\subsection{Track Selection Efficiency}
\label{sec:effic}

\begin{figure}[tbp]
 \begin{center}
  \includegraphics[width=0.95\hsize]{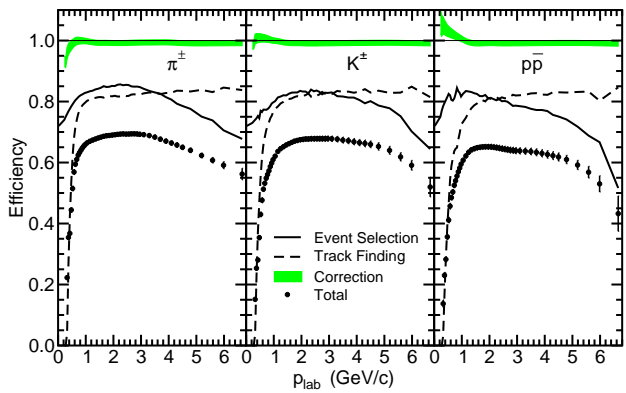}
  \caption{ 
   \label{fig:effevttrk}
   Efficiencies for charged pions (left), kaons (center) and protons
   (right) to produce a selected track in the most forward \cthlab region 
   in the on-resonance sample.
   The solid (dashed) lines represent the simulated fractions of such
   particles that are in selected events
   (of those particles that produce a selected track),
   and the gray/green bands are the products of the corrections
   discussed in the text, 
   with their half-widths representing the uncertainties.
   The points are the products of the three lines/bands.
   }
 \end{center}
\end{figure}

Next, 
we correct the background-subtracted cross sections for the track and
event selection efficiencies,
to obtain corrected cross sections, per hadronic event, for each
hadron type.
Figure~\ref{fig:effevttrk} shows these efficiencies for the three
particle types as functions of \plab in region $\theta 6$.

The solid lines in Fig.~\ref{fig:effevttrk} represent the simulated 
fractions of particles within this \cthlab region that are in a
selected event.
They are well below unity here,
since $\theta 6$ is near the edge of our acceptance,
but peak near 95\% in the central regions.
Most of the \plab dependence arises from the two-jet topology of
\eeqqb events.
Softer tracks are farther, on average, from the thrust axis,
and their distribution becomes nearly isotropic at very low \plab,
where all three fractions approach the average event selection
efficiency of 72\%. 
The highest-\plab tracks tend to define the thrust axis, 
and the fractions drop at high \plab in $\theta 1$ and $\theta 6$,
which span our thrust axis requirement.
The track multiplicity and electron veto criteria introduce
smaller biases against high-\plab tracks in all \cthlab regions.

These biases depend on several aspects of the hadronization process,
which is the object of this measurement.
Since it is not understood in detail,
especially in extreme cases such as events with one very high-momentum 
track,
we compare a number of inclusive track momentum and polar
angle distributions in the data and simulation.
We find a number of inconsistencies, 
some of which are described in Sec.~\ref{sec:results}.
We address these by reweighting the simulated distributions to match
the data,
and by comparing a number of different generators and parameter values
without detector simulation.
We find changes in the event selection bias that are much smaller than
the other uncertainties.

The dashed lines in Fig.~\ref{fig:effevttrk} represent the simulated 
efficiencies for a particle in a selected event to produce a selected
track.
They are zero by definition for tracks with $p_t$ below 0.2~\gevc.
The \pipm efficiency rises rapidly to 80\% at 1~\gevc, 
then increases slowly to an asymptotic value of about 85\%.
The kaon and proton efficiencies rise more slowly due to decays in
flight and interactions in the detector material, respectively,
then show behavior similar to the pions.
This strong similarity is present for prompt particles,
but the pion and proton efficiencies would decrease by up to 10\% 
if \KS and strange baryon decay products were included.
These efficiencies vary little with polar angle.

We perform a number of studies to check and correct the simulated
efficiencies~\cite{tkeff}.
A study of high quality tracks reconstructed in the SVT alone and
extrapolated into the DCH gives information on both the intrinsic
efficiency of the DCH and losses in the material between them.
The simulation is found to be consistent with the data at high \plab,
but corrections of up to 3\% are needed at lower \plab.
A similar study uses pairs of DCH tracks that form a \KS or
$\Lambda$ candidate with a reconstructed vertex inconsistent with
the event vertex but within the innermost layer of the SVT.
This gives information on the SVT efficiency,
indicating the need for 1-3\% corrections at low \plab and 0.5\%
corrections overall.
A further study~\cite{schrist} of identified tracks with a kink 
(which revealed the problem in GEANT4 noted in Sec.~\ref{sec:bkg})
provides a check of the simulation of decays in flight, 
and indicates different material interaction corrections for pions,
kaons and protons.

We also compare the fraction of tracks in the data and simulation that
satisfy each of the selection criteria
after all combinations of the other criteria have been applied.
An overall difference could arise from a deficiency in either the
physics or the detector simulation,
so is of limited use.
However, by studying differences as a function of identified track
type, charge and polar angle, a number of potential
problems with the detector simulation can be corrected or limited.
We find consistency overall, 
but we confirm the discrepancies found in the studies just described,
and also identify a problem with the simulation of the DCH hit
thresholds that affects particles with small \dEdx.
The effect is small except at \plab values near the minimum of the
\dEdx curve in the most central \cthlab region,
where it is as large as 1.3\%.

Combining this information, 
we derive a set of corrections to the simulated efficiencies.
These are shown as bands in Fig.~\ref{fig:effevttrk}, 
where the half-widths indicate the total uncertainties.
The corrections are below 1\% with uncertainties of 0.8\% for 
$\plab \!>\! 1$~\gevc.
At lower momenta, the correction to the kaon efficiency is at the
percent level,
but the pion (proton) efficiency is reduced (increased) by as much as
6.5\% (9\%) with uncertainties of up to one-third of the correction.
The corrections have the same form in the other \cthlab regions, 
but are smaller in proportion to the amount of material traversed.

The simulated interaction rates are different for positively and
negatively charged particles,
as are some of the corrections.
We perform the analysis separately for the two charges up to this
point,
and compare their cross sections at each stage in each \cthlab
region.
Without the efficiency correction, 
we observe differences consistent with expectations.
The fully corrected cross sections are consistent with each other
within the relevant systematic uncertainties.

\subsection{Cross Sections in the CM Frame}
\label{sec:xscm}

At this point we have cross sections for hadrons produced in six
\cthlab regions as functions of their measured \plab.
The measured \plab value can differ from the true value because of
finite momentum resolution,
and low-\plab particles can suffer energy loss before the DCH that
reduces the measured \plab.
These are both small effects on this measurement,
and it is convenient to include corrections for them in the
transformation to the CM frame, discussed in this section.
We verify the quality of our simulation by comparing the masses and
widths of the $\KS \!\to\! \pip\pim$, $\phi \!\to\! \Kp\Km$, 
$\Lambda \!\to\! p\pim$ and $\Lambda_c^+ \!\to\! p\Km\pip$ signals
with those measured in the data (see Ref.~\cite{bbrLcmass}).
The small differences have negligible effects on this measurement,
and we assign no systematic uncertainty from this source.

Differential cross sections in the laboratory frame can, 
in principle, 
be transformed into the CM frame in a model-independent way.
However, 
large nonlinearities for low-\plab and \pstar particles make this
challenging,
and we choose instead a method that is explicitly model dependent,
but allows us to check the simulation at each stage and evaluate
systematic uncertainties reliably.

For each particle type and each \cthlab region,
we first calculate production fractions $F_j$ and an inverse
migration matrix {\boldmath $W$} from the simulation, 
where:
$F_j$ is the fraction of particles produced in the $j^{th}$ \pstar bin
that are boosted into this \cthlab region;
and $W_{ij}$ is the fraction of those boosted into the $i^{th}$ \plab
bin that arise from the $j^{th}$ \pstar bin.
The transformation can then be written as
\begin{equation}
               \mathrm{d} n_j / \mathrm{d} \pstar = (1/F_j)
\sum_i W_{ij} \mathrm{d} n_i / \mathrm{d} \plab.
\end{equation}
The widths of the \cthlab regions are such that
4--5 \plab bins contribute to each \pstar bin at most momenta.
At low $\pstar /m_{had}$, 
this increases to as many as 9 bins for 0.3~\gevc
protons in $\theta 6$.

The matrix {\boldmath $W$} is sensitive to the shape of the true
\pstar distribution,
and the transformation is incorrect if that is not modeled well.
The effect is small (zero) if the distribution varies smoothly
(linearly) over the relevant \pstar range,
but can be large near a peak or inflection point
and at high \pstar where distributions fall exponentially.
We use an iterative procedure
in which we reweight the simulation to match the measured distribution,
redo the transformation,
and repeat until the changes are sufficiently small.
This procedure is reliable if the initial differences are not too
large.
In each case we find a measured \pstar distribution with statistically
significant differences in shape from the simulation,
but the first iteration produces changes smaller than the statistical
uncertainties, 
and changes from the second iteration are negligible.
We assign no systematic uncertainty from this source.

\begin{figure}[tbp]
 \begin{center}
  \includegraphics[width=\hsize]{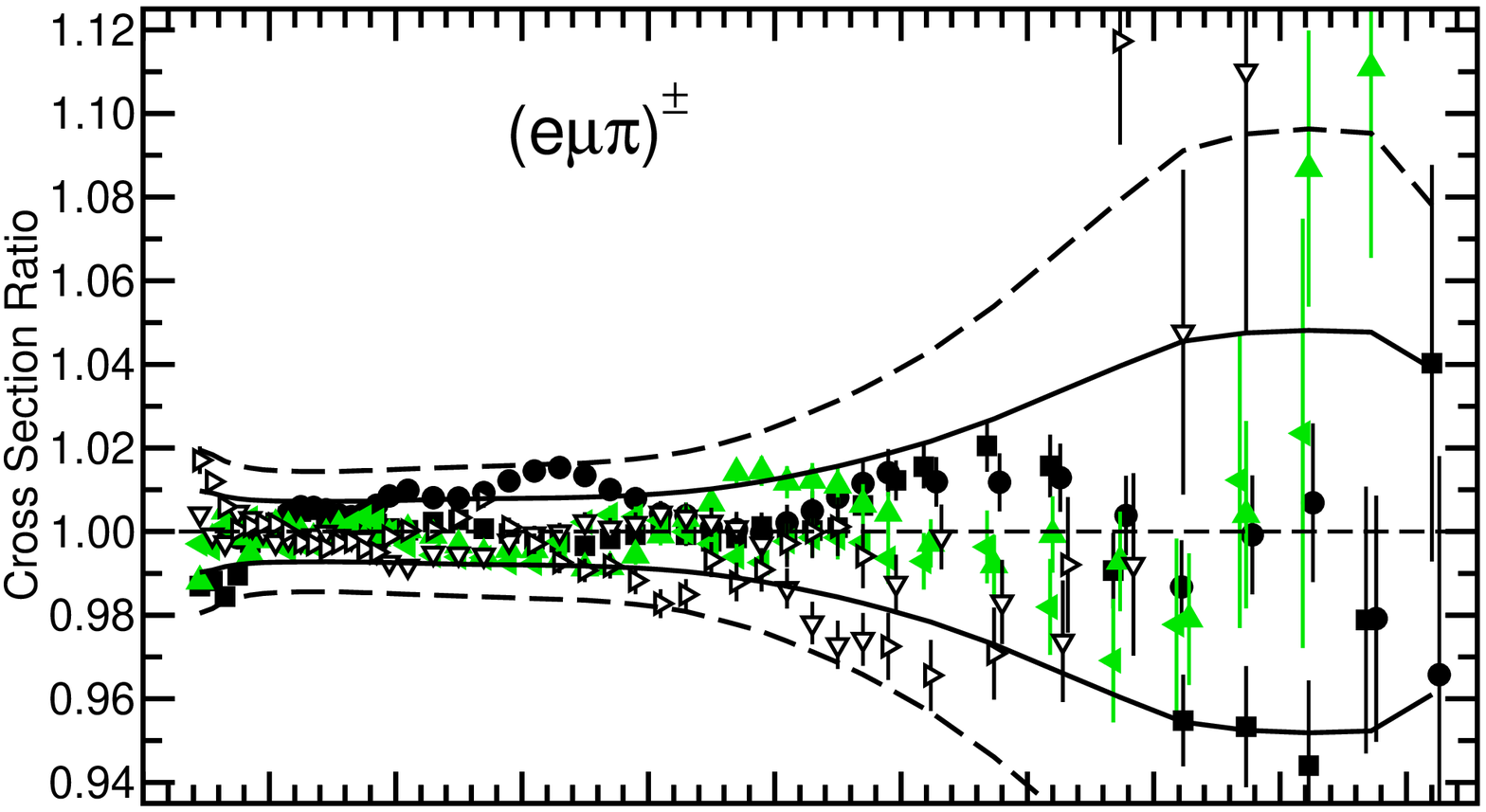}
  \includegraphics[width=\hsize]{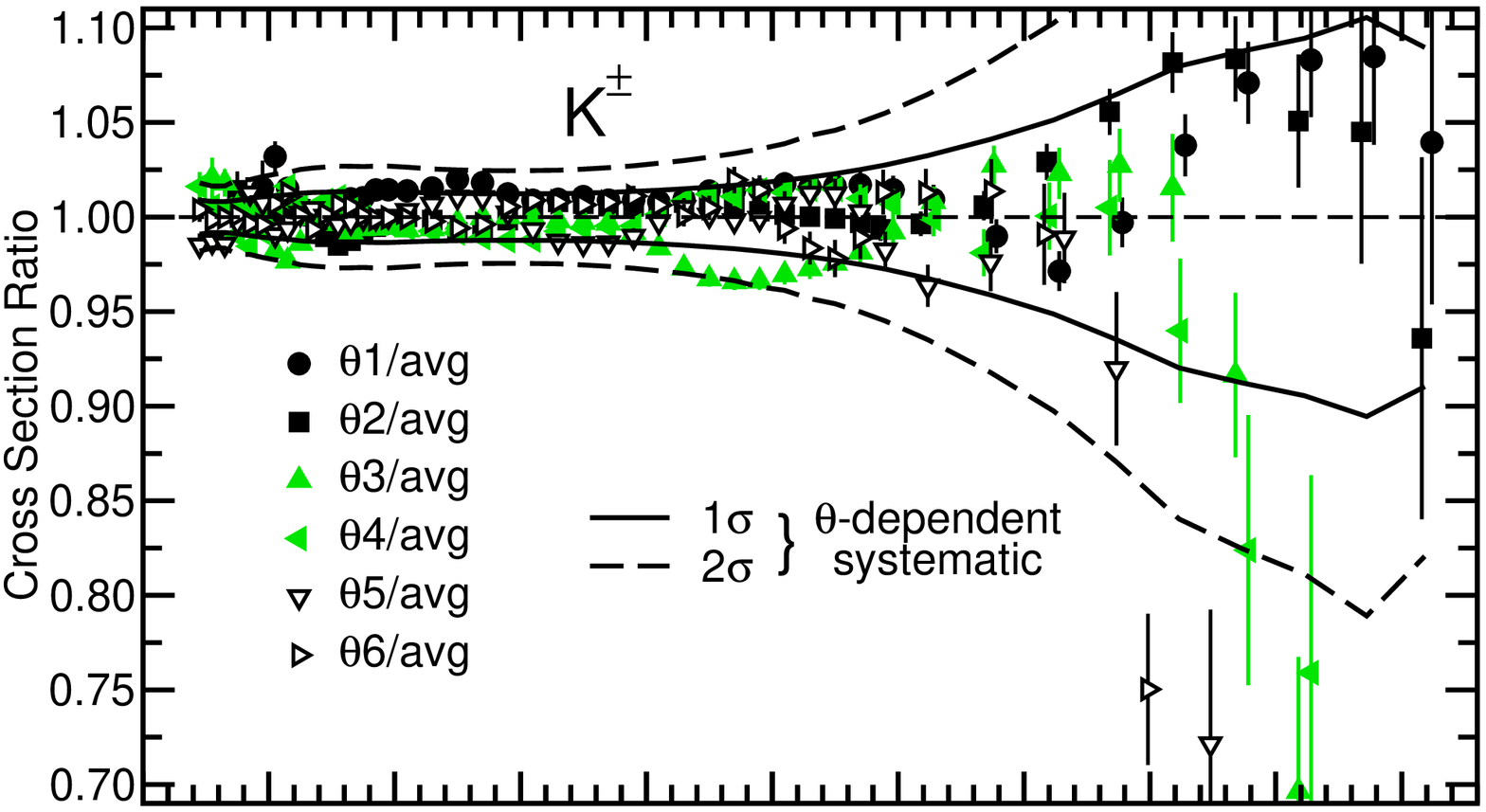}
  \includegraphics[width=\hsize]{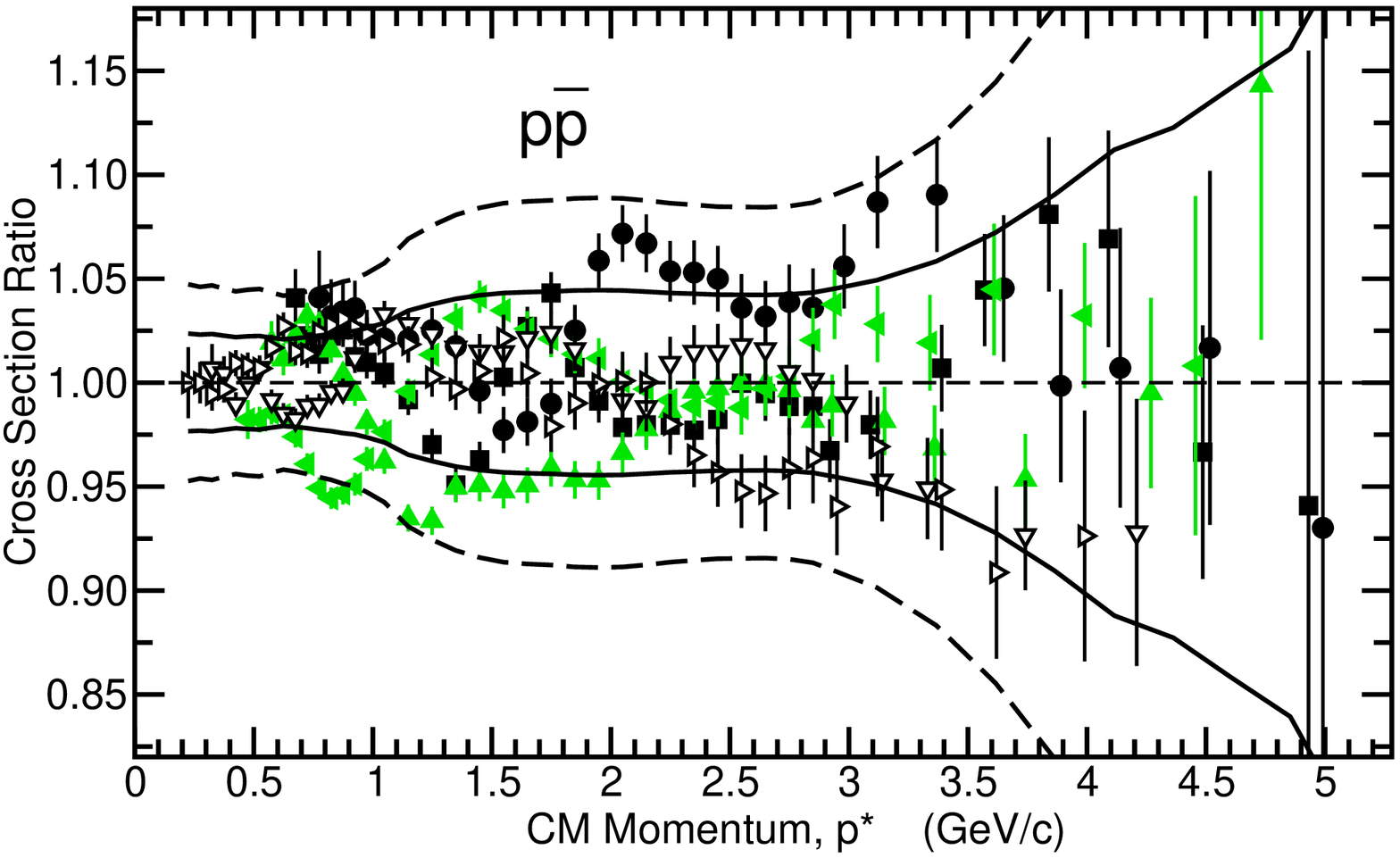}
  \caption{ \label{fig:xsrvcth} 
   Ratios of fully corrected cross sections from each \cthlab region
   to their average value.
   The error bars are statistical only, and some bins have been
   combined for clarity.
   The solid and dashed lines indicate the variations expected at
   one and two standard deviations, respectively.
   }
 \end{center}
\end{figure}

{\boldmath $W$} is insensitive to the true \cthstr distribution,
whereas the $F_j$ are quite sensitive to \cthstr
but almost insensitive to the true \pstar distribution.
The \cthstr distribution must therefore be modeled sufficiently
well.
It has the approximate form 
$D(\cthstr)\propto 1+a(\pstar) \cos^2\thstr$ with $0<a(\pstar)<1$,
where $a(\pstar)$ is small for low \pstar but approaches 1 for
high \pstar.
We compare the cross sections $\sigma_i$ measured in the six \cthlab
regions,
which are shown for the on-resonance data in Fig.~\ref{fig:xsrvcth} 
divided by their weighted average value 
(see below)
in each \pstar bin.
The uncertainties are statistical only, 
and are correlated with their 2--3 nearest neighbors as a result of
the transformation to the CM frame.
The solid (dashed) lines indicate (twice) the root-mean-square (RMS)
variation expected from the systematic uncertainties discussed so
far.
The largest contribution to this is from the particle identification
efficiencies,
which are evaluated independently in each \cthlab region,
but are correlated over ranges of several \pstar bins.
Below 0.5~\gevc,
uncertainties of the tracking efficiency are also important;
these are similar among \cthlab regions for a given \plab value, 
but vary at a given \pstar, 
and are correlated within each \cthlab region.

Overall, the data are consistent within the expected variation,
and the off-resonance data show a similar set of variations.
An incorrectly simulated $a(\pstar)$ would be visible here as a
specific pattern of differences between the $\sigma_i$ at that \pstar, 
roughly parabolic in $i$ with 
$\sigma_1 > \sigma_2 \approx \sigma_6 > \sigma_3 \approx \sigma_5 > \sigma_4$
(or the reverse).
The amplitude of this pattern
would be expected to vary slowly with \pstar over several bins, 
or perhaps across the full range.
No such pattern is visible in Fig.~\ref{fig:xsrvcth},
and we set limits on any mismodeling by fitting the expected pattern
to the $\sigma_i$ in each \pstar bin.
For each particle we find the largest amplitude averaged over three
neighboring \pstar bins.
They correspond to 0.5\%, 1\% and 2\% shifts in $\sigma_1$,
or 1\%, 2\% and 4\% spreads between the $\theta 1$ and $\theta 4$ points
in Fig.~\ref{fig:xsrvcth},
for pions, kaons and protons, respectively.
We take these limiting shifts in $\sigma_1$ as conservative systematic
uncertainties in each region and at all \pstar.
The corresponding shifts in the other $\sigma_i$ are smaller, 
and those in $\sigma_3$, $\sigma_4$ and $\sigma_5$ are of opposite sign;
we take this correlation into account in the average, leading to a
partial cancellation.

This comparison also limits several other systematic effects.
For example,
it is sensitive to an incorrect boost value, which would give 
$\sigma_1 > \sigma_2 > \sigma_3 > \sigma_4 > \sigma_5 > \sigma_6$
with the differences increasing linearly with \pstar;
the data limit any such effect to a negligible level.
A poor simulation of material interactions or soft-track efficiencies
would appear as a spread in the $\sigma_i$, with a particular ordering,
as \pstar approaches its lowest value.
We observe up to 6\% spreads without the corrections described in
Sec.~\ref{sec:effic},
but no significant spread is visible in Fig.~\ref{fig:xsrvcth}.
In the \Kpm plot, 
the highest-\pstar points for $\theta 3$--$\theta 6$ are low.
This may be due to a systematic effect in the kaon identification
efficiencies,
but the uncertainties for these points are large and they contribute little
to the average.
We check for the characteristic ordering and \pstar dependence
expected from residual mismodeling of the \pstar distribution,
and we observe no significant effects.

In each \pstar bin, 
we average the values from the \cthlab regions weighted by their total
uncertainties.
Due to the low identification efficiencies at high \plab,
some measurements have very large uncertainties,
and we do not use the data above \pstar values of 5.00, 4.75, 4.50, and
4.25~\gevc in $\theta 3$, $\theta 4$, $\theta 5$, and $\theta 6$,
respectively.
The uncertainties are relatively large just below these cutoff points,
so that the high-\pstar measurements are dominated by the backward
regions where the momenta are boosted downward and identification
efficiencies are high.
Low-\pstar protons are boosted very far forward,
and only $\theta 6$ contributes below 0.30~\gevc,
with $\theta 5$, $\theta 4$, $\theta 3$, $\theta 2$, and $\theta 1$ 
starting at 0.30, 0.45, 0.60, 0.70, and 0.75~\gevc.
Three (five) regions contribute to the lowest-\pstar kaon (pion) point,
with the others coming in at 0.25, 0.35, and 0.45~\gevc (0.30~\gevc).
All six regions contribute over most of the \pstar range.

The particle-identification uncertainties in different \cthlab regions
are independent of each other,
since they are derived from distinct control samples.
These and the statistical uncertainties are therefore reduced
according to the number of regions contributing to the average in each
\pstar bin.
The uncertainties due to the \cthstr distributions are common to all
\cthstr regions,
but are anticorrelated between the central and forward/backward
regions,
so they are also reduced accordingly.
We take all other uncertainties to be completely correlated between
the regions and average them,
but the weighting takes advantage of the variations with \plab and/or
\cthlab.

\subsection{\boldmath Cross Sections for Prompt and Conventional Hadrons}
\label{sec:xssubtr}

\begin{figure}[tbp]
 \begin{center}
  \includegraphics[width=\hsize]{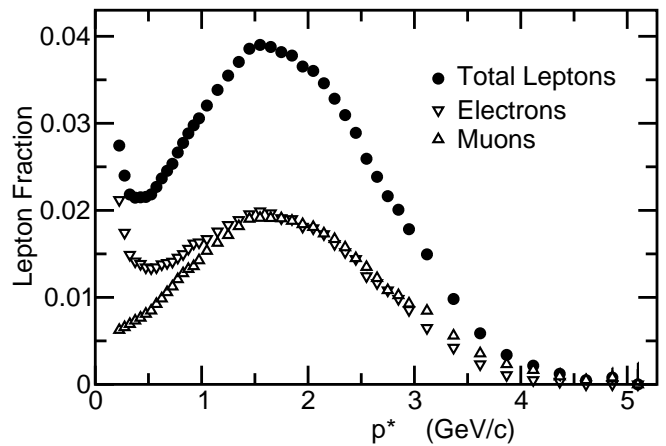}
  \caption{ \label{fig:bkglept}
  Simulated electron (down triangles), muon (up triangles) and total
  lepton (circles) cross sections divided by the \emp cross section
  in \eeqqb events as functions of \pstar.}
 \end{center}
\end{figure}

The leptons in the \emppm cross section are from the
decays of hadrons produced in the fragmentation process, 
such as Dalitz decays of \piz and semileptonic decays of $D$ hadrons.
This 
cross section is included in the supplementary material~\cite{supmat};
we now subtract the leptons to obtain the \pipm cross section.
We show our simulated \epm, \mupm, and total lepton contributions as
fractions of the \emppm cross section in Fig.~\ref{fig:bkglept}.
Charmed hadron decays produce most of the leptons,
with a maximum contribution of 4\% near 1.5~\gevc,
and \piz decays produce most of the \epm at low \pstar.

The \piz cross section has been measured 
in \eeqqb events at higher energies~\cite{pdg},
and the simulation reproduces these results to within 10\%.
Charmed hadron spectra in \eeqqb events have been measured well at
our \ecm~\cite{cleocharm, bellecharm, bbrLcspect}.
Our simulated spectra are slightly too soft, 
which has a small effect on the peak positions in the lepton
spectra.
Of greater concern is the variation in peak position among different
charmed hadrons, 
whose relative production rates are uncertain at the few percent
level~\cite{pdg}.
We subtract the simulated fractional lepton contributions and assign
a set of systematic uncertainties sufficient to cover all these
effects.
We vary the normalization of the \piz contribution by $\pm$10\%,
and consider an independent shape variation by reweighting the \piz
distribution linearly in \pstar so as to change the contribution by
$\pm$50\% at 0.5~\gevc.
We assign a 10\% normalization uncertainty to the charm decay
contribution, 
and also consider a shift in the peak position of $\pm$0.2~\gevc.

\begin{figure}[tbp]
 \begin{center}
  \includegraphics[width=\hsize]{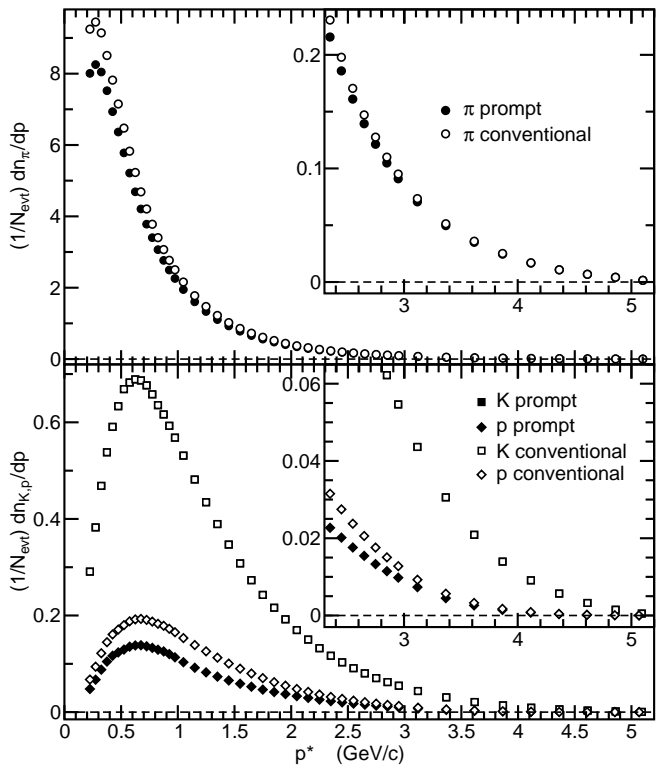}
  \caption{ \label{fig:xsqq} 
   Differential production cross sections for prompt (filled symbols)
   and conventional (open symbols) \pipm (circles), \Kpm (squares) and
   \ppbar (diamonds) per \eeqqb event as functions of \pstar.
   The insets show the high momentum regions with expanded vertical
   scales.
   The prompt and conventional \Kpm cross sections are
   indistinguishable. 
   }
 \end{center}
\end{figure}

The resulting differential cross sections for prompt particles are
shown in Fig.~\ref{fig:xsqq}.
The statistical uncertainties are smaller than the symbol size,
and systematic uncertainties are discussed in the next subsection.
Our measurement covers the bulk of the kaon and proton spectra,
as well as the peak and high side of the pion spectrum.

We calculate cross sections for the conventional set of decay chains
by including the simulated contributions from \KS and strange baryon
decays,
reweighted and with uncertainties as described Sec.~\ref{sec:bkg}.
These conventional cross sections are also shown in 
Fig.~\ref{fig:xsqq}.
The prompt and conventional \Kpm cross sections are indistinguishable,
since the dominant difference is from decays of $\Omega^-$ baryons,
which are produced at a very low rate.
The other cross sections converge at high \pstar
where decays cannot contribute.
The conventional \pipm cross section is a few percent higher overall
than the prompt cross section due to \KS decays,
and as much as 13\% higher at the lowest \pstar due to strange baryon
decays.
The conventional \ppbar cross section is 50\% higher over much of
the range, 
due to strange baryon decays.

\subsection{Summary of Systematic Uncertainties}
\label{sec:systematics}

Most of the systematic uncertainties and checks are described above.
We also consider possible mismodeling of the absolute event selection
efficiency by varying the selection criteria described in
Sec.~\ref{sec:selection} and using alternative event generators.
We find negligible changes in the shapes of the cross sections,
but some of the variations give changes in the overall normalization
of 0.3--0.5\%.
We assign an overall uncertainty of 0.5\%,
corresponding to the largest variation seen.
We also propagate statistical uncertainties on simulated quantities as
a category of systematic uncertainty.

\begin{figure}[tbp]
 \begin{center}
  \includegraphics[width=\hsize]{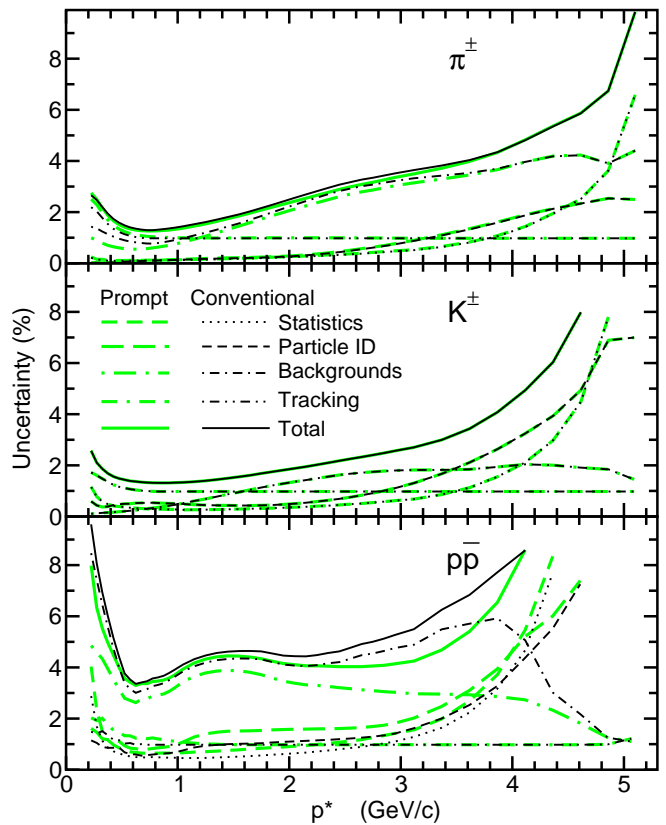}
  \caption{ \label{fig:xserrs} 
   Relative uncertainties in percent on the differential production
   cross sections for prompt (gray/green) and conventional (black)
   \pipm (top), \Kpm (middle) and \ppbar (bottom) 
   as functions of \pstar.
   The totals are shown along with several important components.
   Some uncertainties become large at high \pstar, and are not shown.
   }
 \end{center}
\end{figure}

We summarize the relative uncertainties on our cross sections in
Fig.~\ref{fig:xserrs}.
A few become large at high \pstar,
where the cross sections become low,
and we do not show those that are off the vertical scale.
The statistical uncertainties are much smaller than the systematic
uncertainties except at the highest \pstar values,
and on the lowest-\pstar \ppbar points.
The samples with full detector simulation are similar in size to the
data samples, 
and the corresponding uncertainties (not shown) are the largest
systematic uncertainties at the highest \pstar values.
The total uncertainties are as small as
1.2\%, 1.4\%, and 3.2\% (1.3\%, 1.4\%, and 3.6\% ) for prompt
(conventional) \pipm, \Kpm and \ppbar, respectively, in the
0.6--0.8~\gevc range.
They increase at lower \pstar due mostly to tracking efficiency,
and at higher \pstar due to particle identification and backgrounds.
The latter are dominated by $\tau$-pairs for \pipm and \Kpm,
and by strange decays for \ppbar.

All of the systematic uncertainties have strong point-to-point
correlations.
There is an overall normalization uncertainty of 0.98\% from the event
selection and part of the track-finding efficiency,
which does not affect the shape of any cross section.
The uncertainties due to most backgrounds, strange particle decays, 
\cthstr distributions, and leptons are correlated over wide ranges, 
and can have broad effects on the shape.
Those due to particle identification are correlated strongly over
short ranges, typically $\pm$1--2 neighboring bins, 
and more weakly over $\pm$2--4 additional bins,
and the simulation has been smoothed so that its statistical
uncertainty is correlated over 4--6 bins.
These can lead to apparent structures in the cross sections over ranges
of several bins.
The remaining uncertainties on the tracking efficiencies and 
those due to interactions in the detector material (radiative Bhabha
background) 
are fully correlated over the entire \pstar range,
but are non-negligible only in the 6--10 lowest (highest) \pstar bins.
Overall, the correlation coefficients for neighboring bins are
92--99\% near the centers of the measured ranges and 72--96\%
(15--73\%) toward the low-(high-)\pstar end.
They are over 50\% for bin separations of 12 or fewer.
The full correlation matrices are given in the supplementary 
material~\cite{supmat}.

\section{Results and Interpretation}
\label{sec:results}

\begin{table*}[tpb]
  \caption{\label{tab:dndppmpt}
   Differential cross sections for prompt \pipm, \Kpm and \ppbar in
   \eeqqb events,
   along with their totals over the measured range.
   The first uncertainties are statistical and the second systematic.
   The 0.98\% normalization uncertainty is not included,
   except on the totals.
   }
 \begin{center}
  \begin{tabular}
 {|r@{ -- }l|ll@{$\pm$}l@{$\pm$}l|ll@{$\pm$}l@{$\pm$}l|ll@{$\pm$}l@{$\pm$}l|}
  \hline
   \multicolumn{2}{|c|}{  } &  \multicolumn{4}{c|}{  } &
   \multicolumn{4}{ c|}{  } &  \multicolumn{4}{c|}{  } \\[-0.3cm]
   \multicolumn{2}{|c|}{Momentum} & 
   \multicolumn{4}{ c|}{$(1/N_{\rm evt}) \mbox{d}n_{\pi}/\mbox{d}p$} &
   \multicolumn{4}{ c|}{$(1/N_{\rm evt}) \mbox{d}n_{K}/\mbox{d}p$} &
   \multicolumn{4}{ c|}{$(1/N_{\rm evt}) \mbox{d}n_{p}/\mbox{d}p$} \\[-0.04cm]
   \multicolumn{2}{|c|}{Range (\gevc)} &&
   value & stat. & syst. && Value & stat. & syst. &&
   value & stat. & syst. \\[0.03cm]
  \hline
   \multicolumn{2}{|c|}{  } & \multicolumn{4}{c|}{  } &
   \multicolumn{4}{ c|}{  } & \multicolumn{4}{c|}{  } \\[-0.3cm]
 \hspace*{0.2cm}
 0.20 & 0.25 && 8.01    & 0.02    & 0.22    && 0.291   & 0.003   & 0.007
             && 0.0479  & 0.0019  & 0.0033  \\
 0.25 & 0.30 && 8.25    & 0.01    & 0.21    && 0.382   & 0.003   & 0.007
             && 0.0672  & 0.0016  & 0.0040  \\
 0.30 & 0.35 && 8.04    & 0.01    & 0.17    && 0.468   & 0.003   & 0.008
             && 0.0879  & 0.0013  & 0.0047  \\
 0.35 & 0.40 && 7.52    & 0.01    & 0.14    && 0.538   & 0.003   & 0.009
             && 0.1046  & 0.0012  & 0.0050  \\
 0.40 & 0.45 && 6.93    & 0.01    & 0.11    && 0.591   & 0.002   & 0.009
             && 0.1167  & 0.0011  & 0.0051  \\
 0.45 & 0.50 && 6.36    & 0.01    & 0.10    && 0.633   & 0.002   & 0.009
             && 0.1237  & 0.0011  & 0.0048  \\
 0.50 & 0.55 && 5.78    & 0.01    & 0.08    && 0.669   & 0.002   & 0.009
             && 0.1296  & 0.0010  & 0.0045  \\
 0.55 & 0.60 && 5.21    & 0.01    & 0.07    && 0.682   & 0.002   & 0.009
             && 0.1356  & 0.0009  & 0.0045  \\
 0.60 & 0.65 && 4.69    & 0.01    & 0.06    && 0.689   & 0.002   & 0.009
             && 0.1380  & 0.0009  & 0.0044  \\
 0.65 & 0.70 && 4.21    & 0.01    & 0.05    && 0.687   & 0.002   & 0.009
             && 0.1384  & 0.0009  & 0.0046  \\
 0.70 & 0.75 && 3.781   & 0.005   & 0.048   && 0.676   & 0.002   & 0.009
             && 0.1363  & 0.0009  & 0.0045  \\
 0.75 & 0.80 && 3.402   & 0.004   & 0.043   && 0.658   & 0.002   & 0.008
             && 0.1331  & 0.0009  & 0.0045  \\
 0.80 & 0.85 && 3.065   & 0.004   & 0.039   && 0.636   & 0.002   & 0.008
             && 0.1292  & 0.0008  & 0.0044  \\
 0.85 & 0.90 && 2.765   & 0.004   & 0.035   && 0.616   & 0.002   & 0.008
             && 0.1256  & 0.0008  & 0.0043  \\
 0.90 & 0.95 && 2.495   & 0.003   & 0.032   && 0.593   & 0.002   & 0.008
             && 0.1200  & 0.0008  & 0.0042  \\
 0.95 & 1.00 && 2.258   & 0.003   & 0.030   && 0.568   & 0.002   & 0.007
             && 0.1135  & 0.0008  & 0.0041  \\
 1.00 & 1.10 && 1.948   & 0.002   & 0.027   && 0.531   & 0.001   & 0.007
             && 0.1033  & 0.0007  & 0.0040  \\
 1.10 & 1.20 && 1.603   & 0.002   & 0.023   && 0.482   & 0.001   & 0.006
             && 0.0919  & 0.0006  & 0.0038  \\
 1.20 & 1.30 && 1.332   & 0.002   & 0.020   && 0.435   & 0.001   & 0.006
             && 0.0823  & 0.0006  & 0.0035  \\
 1.30 & 1.40 && 1.106   & 0.002   & 0.018   && 0.389   & 0.001   & 0.006
             && 0.0734  & 0.0005  & 0.0032  \\
 1.40 & 1.50 && 0.926   & 0.002   & 0.016   && 0.347   & 0.001   & 0.005
             && 0.0655  & 0.0005  & 0.0029  \\
 1.50 & 1.60 && 0.780   & 0.002   & 0.014   && 0.3080  & 0.0009  & 0.0047
             && 0.0588  & 0.0005  & 0.0026  \\
 1.60 & 1.70 && 0.659   & 0.001   & 0.013   && 0.2731  & 0.0008  & 0.0043
             && 0.0526  & 0.0004  & 0.0023  \\
 1.70 & 1.80 && 0.559   & 0.001   & 0.012   && 0.2427  & 0.0008  & 0.0040
             && 0.0466  & 0.0004  & 0.0020  \\
 1.80 & 1.90 && 0.475   & 0.001   & 0.010   && 0.2161  & 0.0007  & 0.0037
             && 0.0416  & 0.0004  & 0.0017  \\
 1.90 & 2.00 && 0.404   & 0.001   & 0.009   && 0.1921  & 0.0007  & 0.0034
             && 0.0374  & 0.0003  & 0.0015  \\
 2.00 & 2.10 && 0.343   & 0.001   & 0.008   && 0.1698  & 0.0006  & 0.0031
             && 0.0331  & 0.0003  & 0.0013  \\
 2.10 & 2.20 && 0.294   & 0.001   & 0.007   && 0.1503  & 0.0006  & 0.0029
             && 0.0293  & 0.0003  & 0.0012  \\
 2.20 & 2.30 && 0.251   & 0.001   & 0.007   && 0.1323  & 0.0005  & 0.0026
             && 0.0259  & 0.0003  & 0.0010  \\
 2.30 & 2.40 && 0.216   & 0.001   & 0.006   && 0.1167  & 0.0005  & 0.0024
             && 0.0227  & 0.0002  & 0.0009  \\
 2.40 & 2.50 && 0.186   & 0.001   & 0.005   && 0.1031  & 0.0005  & 0.0022
             && 0.0201  & 0.0002  & 0.0008  \\
 2.50 & 2.60 && 0.1610  & 0.0006  & 0.0048  && 0.0909  & 0.0004  & 0.0020
             && 0.0176  & 0.0002  & 0.0007  \\
 2.60 & 2.70 && 0.1394  & 0.0005  & 0.0043  && 0.0802  & 0.0004  & 0.0018
             && 0.0154  & 0.0002  & 0.0006  \\
 2.70 & 2.80 && 0.1213  & 0.0005  & 0.0038  && 0.0704  & 0.0004  & 0.0016
             && 0.0133  & 0.0002  & 0.0005  \\
 2.80 & 2.90 && 0.1048  & 0.0005  & 0.0034  && 0.0622  & 0.0004  & 0.0015
             && 0.01146 & 0.00015 & 0.00044 \\
 2.90 & 3.00 && 0.0910  & 0.0004  & 0.0030  && 0.0546  & 0.0003  & 0.0014
             && 0.00979 & 0.00014 & 0.00038 \\
 3.00 & 3.25 && 0.0706  & 0.0004  & 0.0024  && 0.0436  & 0.0003  & 0.0011 
             && 0.00733 & 0.00011 & 0.00029 \\
 3.25 & 3.50 && 0.0497  & 0.0003  & 0.0018  && 0.0306  & 0.0003  & 0.0009
             && 0.00448 & 0.00009 & 0.00019 \\
 3.50 & 3.75 && 0.0350  & 0.0003  & 0.0014  && 0.0209  & 0.0002  & 0.0007
             && 0.00260 & 0.00007 & 0.00012 \\
 3.75 & 4.00 && 0.0246  & 0.0003  & 0.0010  && 0.0139  & 0.0002  & 0.0005
             && 0.00143 & 0.00005 & 0.00008 \\
 4.00 & 4.25 && 0.0167  & 0.0002  & 0.0008  && 0.00910 & 0.00019 & 0.00041
             && 0.00073 & 0.00004 & 0.00005 \\
 4.25 & 4.50 && 0.0107  & 0.0002  & 0.0005  && 0.00568 & 0.00017 & 0.00030
             && 0.00036 & 0.00003 & 0.00003 \\
 4.50 & 4.75 && 0.00681 & 0.00017 & 0.00036 && 0.00324 & 0.00015 & 0.00021
             && 0.00017 & 0.00002 & 0.00002 \\
 4.75 & 5.00 && 0.00418 & 0.00015 & 0.00024 && 0.00149 & 0.00012 & 0.00015
             && 0.00007 & 0.00002 & 0.00001 \\
 5.00 & 5.27 && 0.00153 & 0.00010 & 0.00011  \hspace*{0.1cm} 
             && 0.00050 & 0.00007 & 0.00007  \hspace*{0.1cm} 
             && 0.00001 & 0.00001 & 0.00001  \hspace*{0.1cm} \\[ 0.01cm]
  \hline
   \multicolumn{2}{|c|}{  } & \multicolumn{4}{c|}{  } &
   \multicolumn{4}{ c|}{  } & \multicolumn{4}{c|}{  } \\[-0.31cm]
 0.20 & 5.27 && 5.364   & 0.002   & 0.080   && 0.946    & 0.001   & 0.012
             && 0.1819  & 0.0003  & 0.0058  \\[ 0.01cm]
  \hline
  \end{tabular}
 \end{center}
\end{table*}

\begin{table*}[tpb]
 \begin{center}
  \caption{\label{tab:dndpconv}
   Differential cross sections for conventional \pipm, \Kpm and \ppbar in
   \eeqqb events,
   along with their totals over the measured range.
   The first uncertainties are statistical and the second systematic.
   The 0.98\% normalization uncertainty is not included,
   except on the totals.
   }
  \begin{tabular}
 {|r@{ -- }l|ll@{$\pm$}l@{$\pm$}l|ll@{$\pm$}l@{$\pm$}l|ll@{$\pm$}l@{$\pm$}l|}
  \hline
   \multicolumn{2}{|c|}{  } &  \multicolumn{4}{c|}{  } &
   \multicolumn{4}{ c|}{  } &  \multicolumn{4}{c|}{  } \\[-0.3cm]
   \multicolumn{2}{|c|}{Momentum} & 
   \multicolumn{4}{ c|}{$(1/N_{\rm evt}) \mbox{d}n_{\pi}/\mbox{d}p$} &
   \multicolumn{4}{ c|}{$(1/N_{\rm evt}) \mbox{d}n_{K}/\mbox{d}p$} &
   \multicolumn{4}{ c|}{$(1/N_{\rm evt}) \mbox{d}n_{p}/\mbox{d}p$} \\[-0.04cm]
   \multicolumn{2}{|c|}{Range (\gevc)} &&
   value & stat. & syst. && Value & stat. & syst. &&
   value & stat. & syst. \\[0.03cm]
  \hline
   \multicolumn{2}{|c|}{  } & \multicolumn{4}{c|}{  } &
   \multicolumn{4}{ c|}{  } & \multicolumn{4}{c|}{  } \\[-0.3cm]
 \hspace*{0.2cm}
 0.20 & 0.25 && 9.25    & 0.02    & 0.24    && 0.291   & 0.003   & 0.007
             && 0.068   & 0.002   & 0.006   \\
 0.25 & 0.30 && 9.45    & 0.01    & 0.23    && 0.383   & 0.003   & 0.007
             && 0.094   & 0.002   & 0.007   \\
 0.30 & 0.35 && 9.14    & 0.01    & 0.20    && 0.468   & 0.003   & 0.008
             && 0.122   & 0.001   & 0.008   \\
 0.35 & 0.40 && 8.51    & 0.01    & 0.16    && 0.538   & 0.003   & 0.009
             && 0.145   & 0.001   & 0.009   \\
 0.40 & 0.45 && 7.82    & 0.01    & 0.14    && 0.591   & 0.002   & 0.009
             && 0.161   & 0.001   & 0.008   \\
 0.45 & 0.50 && 7.15    & 0.01    & 0.11    && 0.633   & 0.002   & 0.009
             && 0.171   & 0.001   & 0.007   \\
 0.50 & 0.55 && 6.47    & 0.01    & 0.10    && 0.669   & 0.002   & 0.009
             && 0.180   & 0.001   & 0.007   \\
 0.55 & 0.60 && 5.83    & 0.01    & 0.08    && 0.683   & 0.002   & 0.009
             && 0.188   & 0.001   & 0.007   \\
 0.60 & 0.65 && 5.23    & 0.01    & 0.07    && 0.689   & 0.002   & 0.009
             && 0.192   & 0.001   & 0.006   \\
 0.65 & 0.70 && 4.69    & 0.01    & 0.06    && 0.687   & 0.002   & 0.009
             && 0.193   & 0.001   & 0.007   \\
 0.70 & 0.75 && 4.21    & 0.01    & 0.05    && 0.677   & 0.002   & 0.009
             && 0.191   & 0.001   & 0.007   \\
 0.75 & 0.80 && 3.778   & 0.004   & 0.049   && 0.658   & 0.002   & 0.008
             && 0.187   & 0.001   & 0.007   \\
 0.80 & 0.85 && 3.401   & 0.004   & 0.044   && 0.636   & 0.002   & 0.008
             && 0.183   & 0.001   & 0.007   \\
 0.85 & 0.90 && 3.067   & 0.004   & 0.041   && 0.616   & 0.002   & 0.008
             && 0.179   & 0.001   & 0.007   \\
 0.90 & 0.95 && 2.768   & 0.003   & 0.038   && 0.593   & 0.002   & 0.008
             && 0.173   & 0.001   & 0.006   \\
 0.95 & 1.00 && 2.504   & 0.003   & 0.035   && 0.568   & 0.002   & 0.007
             && 0.165   & 0.001   & 0.006   \\
 1.00 & 1.10 && 2.159   & 0.003   & 0.031   && 0.531   & 0.001   & 0.007
             && 0.153   & 0.001   & 0.006  \\
 1.10 & 1.20 && 1.775   & 0.002   & 0.027   && 0.482   & 0.001   & 0.006
             && 0.139   & 0.001   & 0.006  \\
 1.20 & 1.30 && 1.472   & 0.002   & 0.024   && 0.435   & 0.001   & 0.006
             && 0.126   & 0.001   & 0.006  \\
 1.30 & 1.40 && 1.221   & 0.002   & 0.021   && 0.389   & 0.001   & 0.006 
             && 0.113   & 0.001   & 0.005  \\
 1.40 & 1.50 && 1.020   & 0.002   & 0.018   && 0.347   & 0.001   & 0.005
             && 0.1006  & 0.0005  & 0.0046 \\
 1.50 & 1.60 && 0.857   & 0.002   & 0.016   && 0.3080  & 0.0009  & 0.0047
             && 0.0900  & 0.0005  & 0.0042 \\
 1.60 & 1.70 && 0.723   & 0.001   & 0.015   && 0.2731  & 0.0008  & 0.0043
             && 0.0799  & 0.0004  & 0.0037 \\
 1.70 & 1.80 && 0.611   & 0.001   & 0.013   && 0.2427  & 0.0008  & 0.0040
             && 0.0704  & 0.0004  & 0.0032  \\
 1.80 & 1.90 && 0.518   & 0.001   & 0.012   && 0.2161  & 0.0007  & 0.0037
             && 0.0620  & 0.0004  & 0.0028  \\
 1.90 & 2.00 && 0.439   & 0.001   & 0.011   && 0.1921  & 0.0007  & 0.0034
             && 0.0548  & 0.0003  & 0.0024  \\
 2.00 & 2.10 && 0.372   & 0.001   & 0.009   && 0.1698  & 0.0006  & 0.0031
             && 0.0480  & 0.0003  & 0.0021  \\
 2.10 & 2.20 && 0.317   & 0.001   & 0.008   && 0.1503  & 0.0006  & 0.0029
             && 0.0419  & 0.0003  & 0.0018  \\
 2.20 & 2.30 && 0.270   & 0.001   & 0.008   && 0.1323  & 0.0005  & 0.0026
             && 0.0364  & 0.0003  & 0.0016  \\
 2.30 & 2.40 && 0.231   & 0.001   & 0.007   && 0.1167  & 0.0005  & 0.0024
             && 0.0315  & 0.0002  & 0.0014  \\
 2.40 & 2.50 && 0.198   & 0.001   & 0.006   && 0.1031  & 0.0005  & 0.0022
             && 0.0275  & 0.0002  & 0.0012  \\
 2.50 & 2.60 && 0.170   & 0.001   & 0.005   && 0.0909  & 0.0004  & 0.0020
             && 0.0237  & 0.0002  & 0.0011  \\
 2.60 & 2.70 && 0.1471  & 0.0006  & 0.0048  && 0.0802  & 0.0004  & 0.0018
             && 0.0206  & 0.0002  & 0.0010 \\
 2.70 & 2.80 && 0.1276  & 0.0005  & 0.0042  && 0.0704  & 0.0004  & 0.0016
             && 0.0176  & 0.0002  & 0.0009 \\
 2.80 & 2.90 && 0.1099  & 0.0005  & 0.0037  && 0.0622  & 0.0004  & 0.0015
             && 0.0150  & 0.0002  & 0.0008 \\
 2.90 & 3.00 && 0.0950  & 0.0004  & 0.0033  && 0.0546  & 0.0003  & 0.0014 
             && 0.0128  & 0.0001  & 0.0007 \\
 3.00 & 3.25 && 0.0734  & 0.0004  & 0.0026  && 0.0436  & 0.0003  & 0.0011 
             && 0.00926 & 0.00012 & 0.00049 \\
 3.25 & 3.50 && 0.0513  & 0.0003  & 0.0019  && 0.0306  & 0.0003  & 0.0009
             && 0.00564 & 0.00009 & 0.00034 \\
 3.50 & 3.75 && 0.0359  & 0.0003  & 0.0014  && 0.0209  & 0.0002  & 0.0007
             && 0.00324 & 0.00007 & 0.00021 \\
 3.75 & 4.00 && 0.0251  & 0.0003  & 0.0011  && 0.0139  & 0.0002  & 0.0005
             && 0.00173 & 0.00005 & 0.00012 \\
 4.00 & 4.25 && 0.0169  & 0.0002  & 0.0008  && 0.00910 & 0.00019 & 0.00041
             && 0.00087 & 0.00004 & 0.00006 \\
 4.25 & 4.50 && 0.0108  & 0.0002  & 0.0005  && 0.00568 & 0.00017 & 0.00030
             && 0.00040 & 0.00003 & 0.00003 \\
 4.50 & 4.75 && 0.00682 & 0.00017 & 0.00036 && 0.00324 & 0.00015 & 0.00021
             && 0.00017 & 0.00002 & 0.00002 \\
 4.75 & 5.00 && 0.00418 & 0.00015 & 0.00024 && 0.00149 & 0.00012 & 0.00015
             && 0.00007 & 0.00002 & 0.00001 \\
 5.00 & 5.27 && 0.00153 & 0.00010 & 0.00011  \hspace*{0.1cm} 
             && 0.00050 & 0.00007 & 0.00007  \hspace*{0.1cm} 
             && 0.00001 & 0.00001 & 0.00001  \hspace*{0.1cm} \\[ 0.03cm]
  \hline
   \multicolumn{2}{|c|}{  } & \multicolumn{4}{c|}{  } &
   \multicolumn{4}{ c|}{  } & \multicolumn{4}{c|}{  } \\[-0.3cm]
 0.20 & 5.27 && 6.002   & 0.002   & 0.092   && 0.946    & 0.001   & 0.012
             && 0.2612  & 0.0003  & 0.0095  \\[ 0.03cm]
  \hline
  \end{tabular}
 \end{center}
\end{table*}

Our results for prompt and conventional hadrons are listed in
Tables~\ref{tab:dndppmpt} and~\ref{tab:dndpconv}.
Several other tables, 
including breakdowns of the uncertainties and their correlations, 
are available in the supplementary material~\cite{supmat}.
In this section,
we compare the cross section results with previous measurements,
models of hadronization,
and predictions of QCD.
We also calculate average event multiplicities,
ratios of differential production cross sections,
and charged hadron fractions.

\subsection{Cross Sections in \eeqqb Events}
\label{sec:resqqb}

\begin{figure}[tbp]
 \begin{center}
  \includegraphics[width=\hsize]{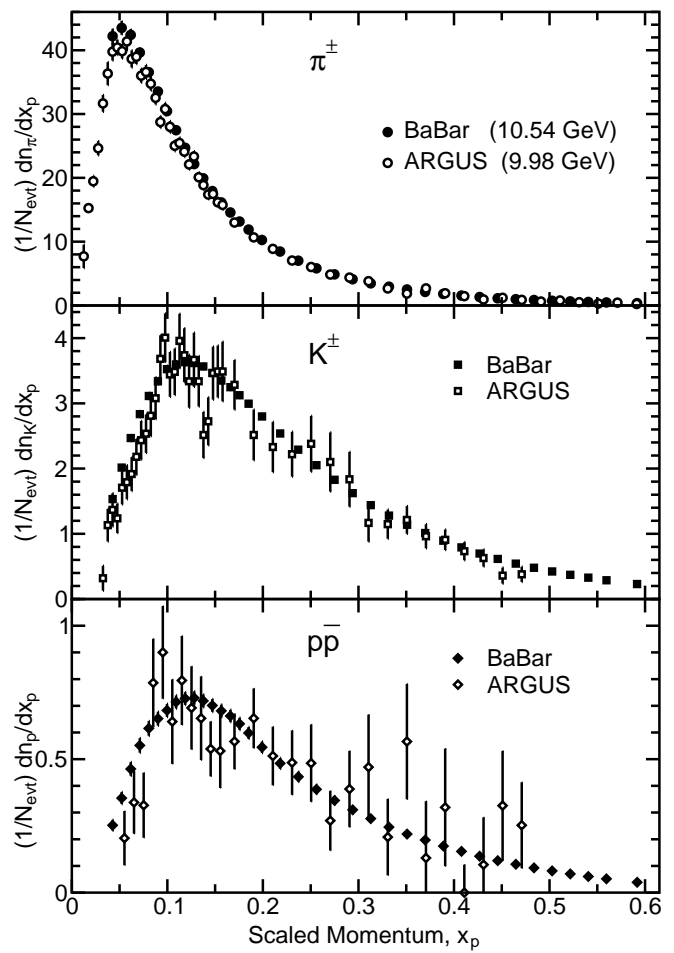}
 \end{center}
 \caption{ \label{fig:xsqqvold} 
  Comparison of our differential cross sections for prompt \pipm (top), 
  \Kpm (middle) and \ppbar (bottom) with previous results from ARGUS at
  $\sqrt{s} \!=\! 9.98$~\gev.
  The error bars represent combined statistical and systematic
  uncertainties.
  }
\end{figure}

We compare our results with previous measurements 
from the ARGUS experiment~\cite{argus89} 
of differential \pipm, \Kpm and \ppbar production cross sections in
\eeqqb events at the slightly lower $\ecm \!=\! 9.98$~\gev.
Figure~\ref{fig:xsqqvold} shows their tabulated results for prompt
particles, along with ours, in terms of the scaled momentum 
$x_p \!=\! 2\pstar / \ecm$, 
over the range of their measurements.
Total uncertainties are shown for both data sets.
Although our results are far more precise statistically,
the systematic uncertainties are comparable and are correlated over
significant $x_p$ ranges in both cases.
The ARGUS \pipm and \Kpm data extend to lower $x_p$ values,
whereas ours extend up to $x_p \!=\! 1$,
so that most of the relevant range is covered between the
two experiments.

For $x_p \! >\! 0.1$, 
the two data sets are consistent.
As $x_p$ decreases, 
the ARGUS data fall systematically below ours,
as might be expected from a mass-driven scaling violation.
The differences are consistent with those expected by the
hadronization models described in Sec.~\ref{sec:resmodels}.
However, 
when the correlations between the systematic uncertainties are taken
into account,
the significance of these differences is only a few standard
deviations for \pipm and \Kpm, and below 2$\sigma$ for \ppbar.
ARGUS also presents results including \KS and $\Lambda$ decay products.
A comparison with our conventional results yields the same conclusions.

\subsection{Comparison with Hadronization Models}
\label{sec:resmodels}

\begin{figure}[tbp]
 \begin{center}
  \includegraphics[width=\hsize]{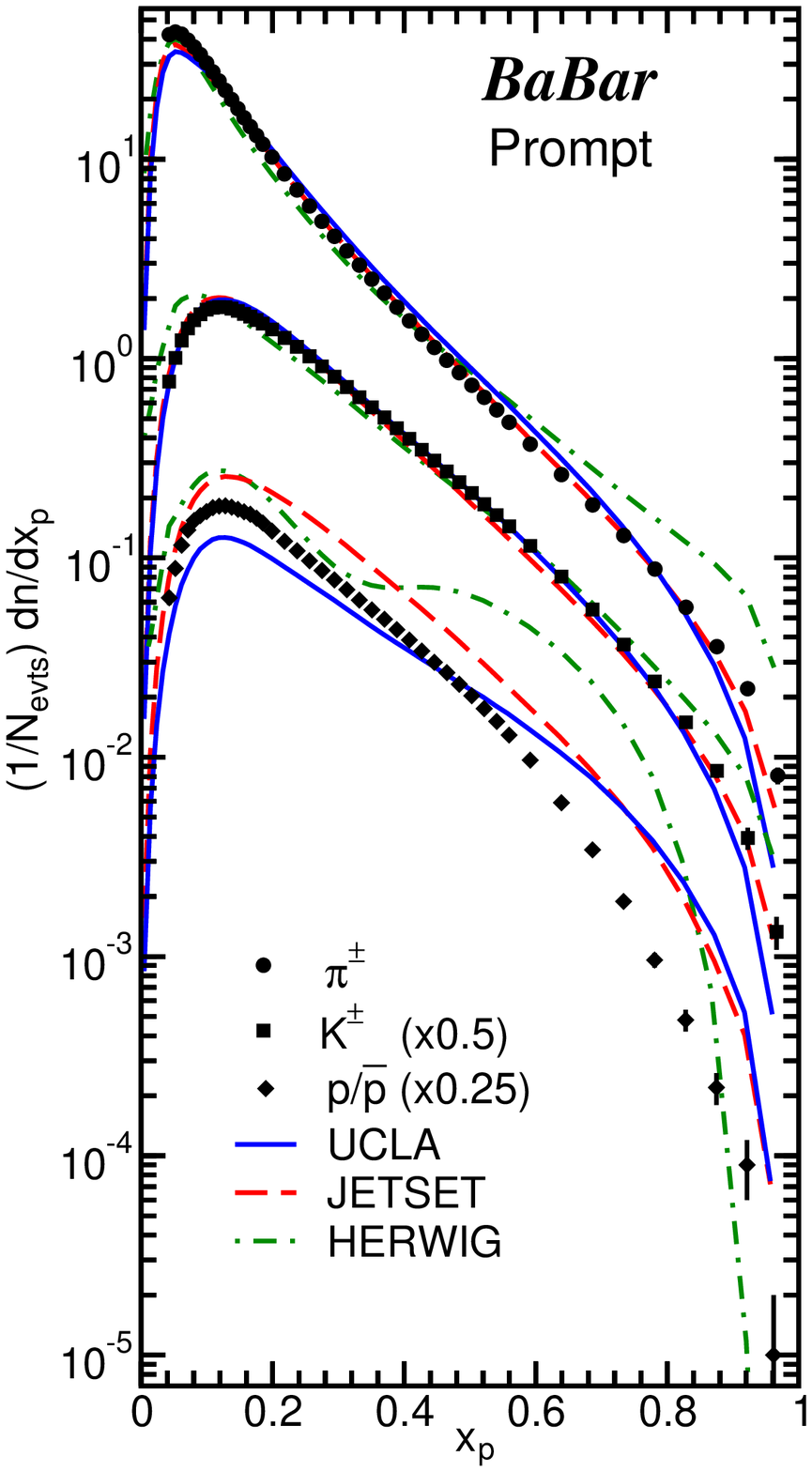}
 \end{center}
 \caption{ \label{fig:xsqqmc} 
  Comparison of the prompt \pipm (circles), \Kpm (squares) and \ppbar
  (diamonds) cross sections in \eeqqb events with the predictions of the
  UCLA (solid line), JETSET (dashed) and HERWIG (dotted) hadronization
  models.
  }
\end{figure}

In Fig.~\ref{fig:xsqqmc},
we compare our cross sections for prompt particles with the
predictions of the three hadronization models discussed in
Sec.~\ref{sec:intro}.
These models represent the three different mechanisms for 
hadronization currently available.
In each case we use the default parameter values,
which have been chosen based on previous data, 
mostly at higher energies but including the ARGUS data.
All three models describe the bulk of the spectra qualitatively,
but no model describes any spectrum in detail.
The peak positions are consistent with the data, 
except for the HERWIG \Kpm, which is too low.
The peak amplitudes are low by 9--20\% for \pipm,
high by 8--11\% for \Kpm,
and either 30\% low or 30--50\% high for \ppbar.

The HERWIG peaks are too narrow, 
and the high-$x_p$ tails are much too long;
in particular, the \ppbar spectrum shows a pronounced structure at
high $x_p$, 
and also drops to zero in the highest-$x_p$ bin.
In contrast, 
the JETSET and UCLA \pipm and \Kpm peaks are slightly too broad, 
and the tails too short,
although both models describe the shape well in the 0.2--0.7 range.
UCLA also reproduces the amplitude of the \Kpm spectrum in this range.
JETSET's \ppbar spectrum has the correct shape for $x_p \!<\! 0.5$,
but then drops too slowly.
UCLA's \ppbar spectrum is distorted relative to the data in a manner
similar to HERWIG's \pipm and \Kpm spectra.
A comparison of the conventional cross sections (not shown) gives
similar results.

Similar discrepancies with these models have been reported at higher
energies~\cite{pikptpc,pikptasso,pikptopaz,pikpdelphi,pikpopal,pikpaleph,pikpsld}, 
although earlier versions of the models were often used and some parameter
values differed.
Most differences from the data were of the same sign and similar in size to
those we observe,
suggesting that the scaling with \ecm might be well simulated.
In some cases,
simple changes to parameters in JETSET produced improvements in the
agreement with data, and some experiments implemented global tuning.
We do not attempt to tune any of the models,
but we test some simple modifications of JETSET parameters:
changing 
the probability of producing a diquark-antidiquark, 
rather than a \qqbar, pair at each string break modifies the amplitude
of the simulated 
proton spectrum, but does not change the shape;
similarly, the probability to produce an \ssbar,
rather than \uubar or \ddbar, pair controls the amplitude, 
but not the shape, of the kaon spectrum.

\begin{figure}[tbp]
 \begin{center}
  \includegraphics[width=\hsize]{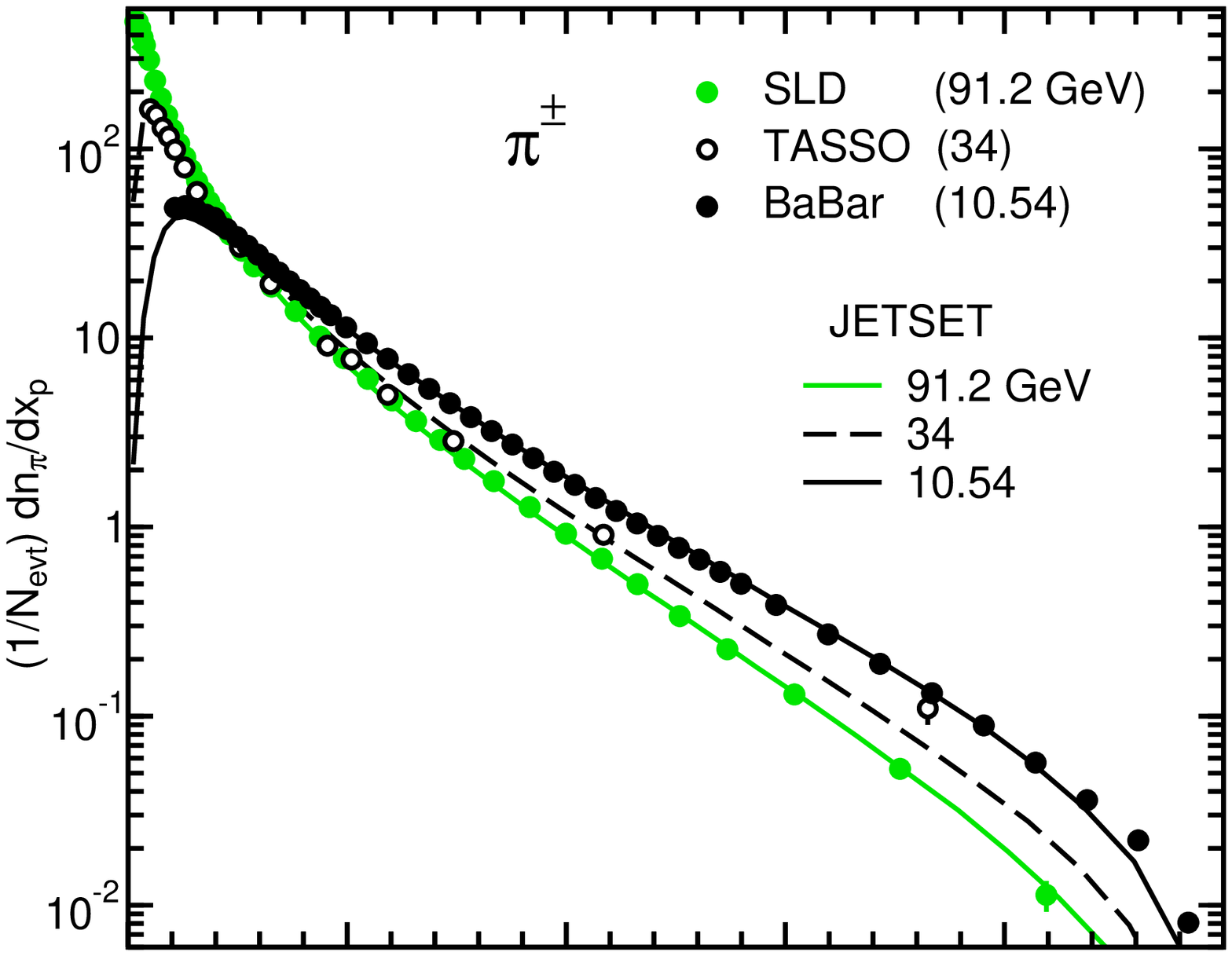}
  \includegraphics[width=\hsize]{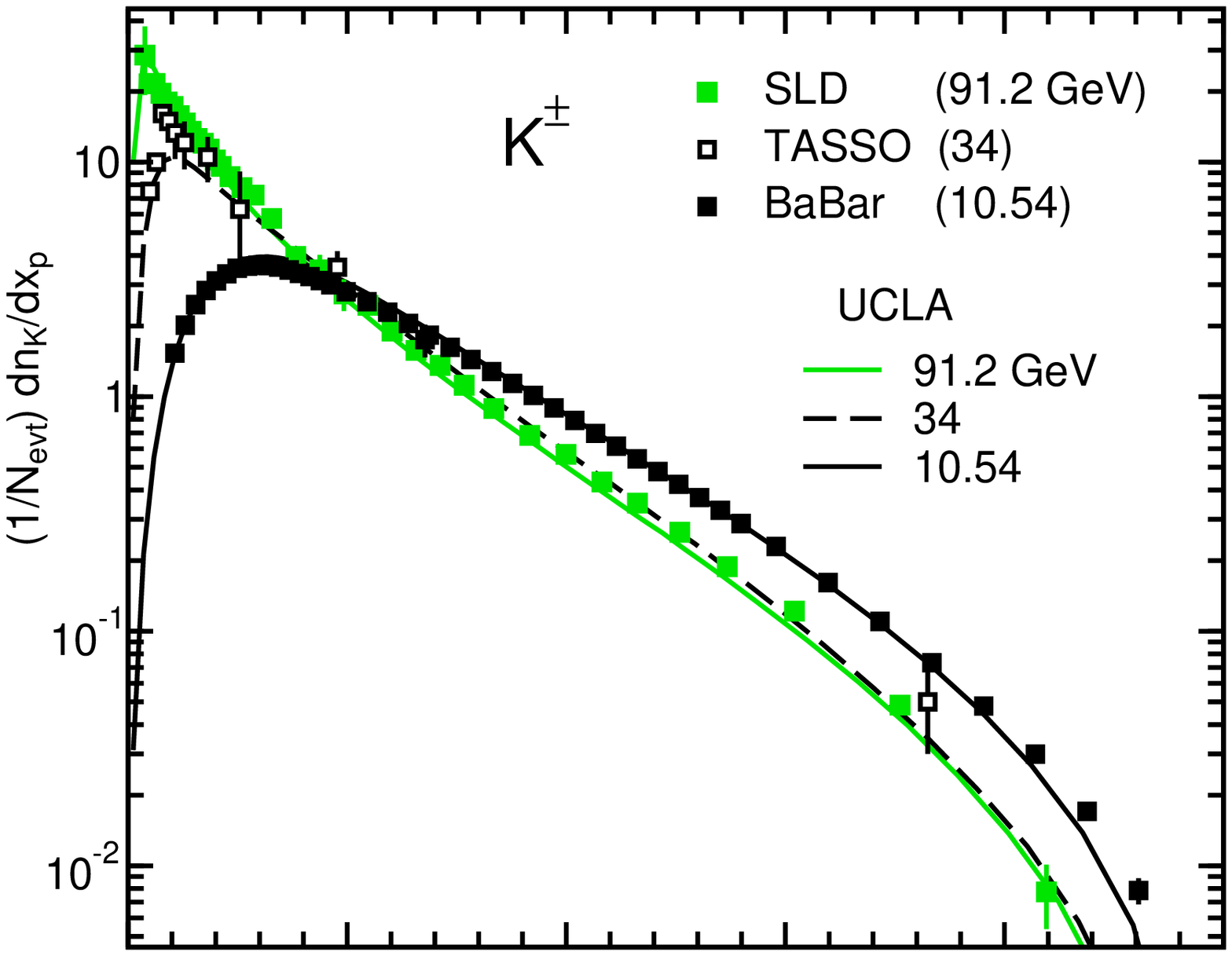}
  \includegraphics[width=\hsize]{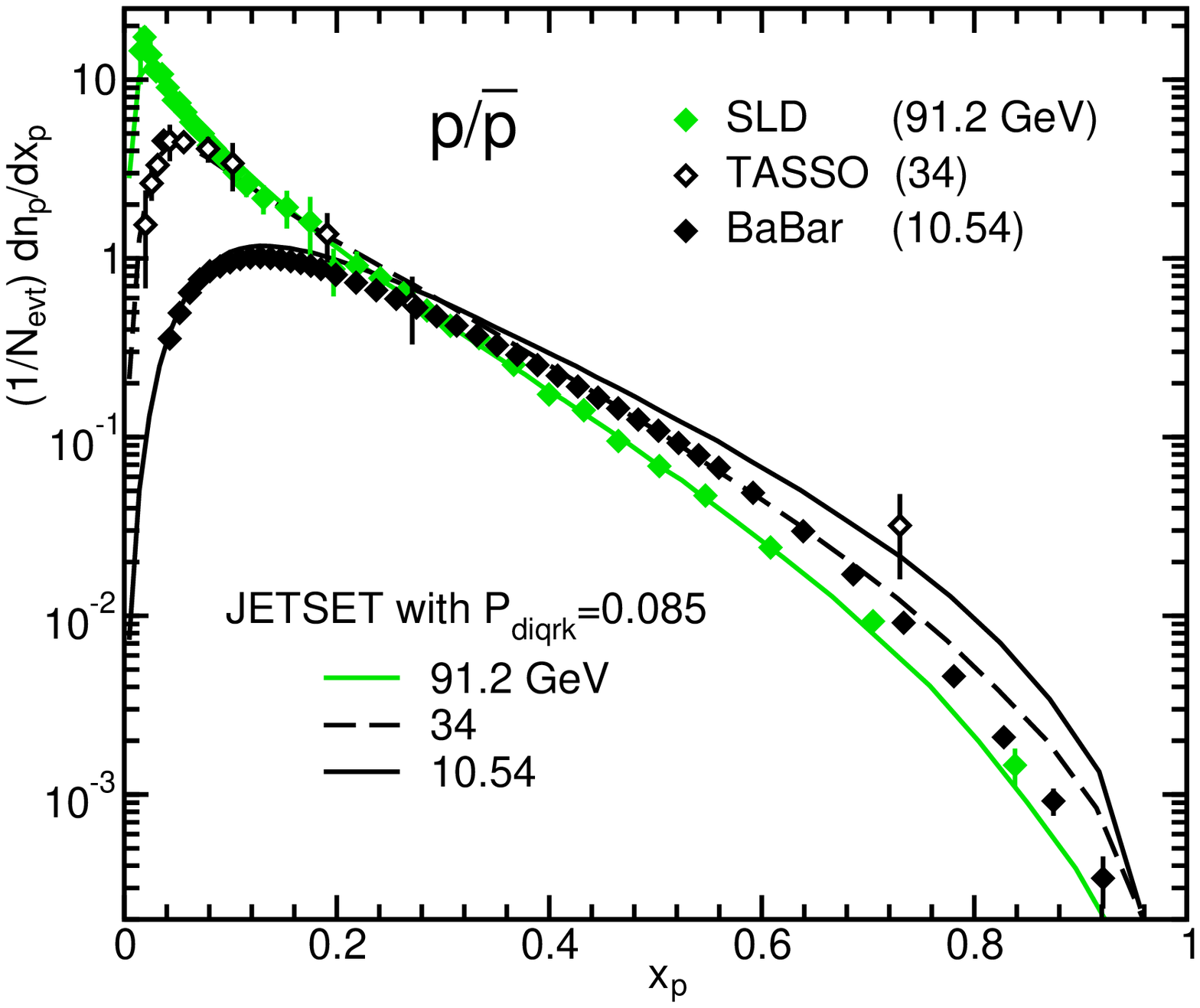}
 \end{center}
 \caption{ \label{fig:xsqqscal} 
   Conventional \pipm (top), \Kpm (middle) and \ppbar (bottom) cross
   sections measured at three different CM energies,
   compared with the predictions of the simulations described in the
   text.
   }
\end{figure}

We test the scaling properties of the models by generating samples
with each at various energies, 
comparing them with available data,
and looking for changes in the type or magnitude of any differences.
In the top plot in Fig.~\ref{fig:xsqqscal} 
we show our conventional \pipm cross
section along with those from the TASSO and SLD experiments.
At high $x_p$,
these two experiments provide the most precise data and/or widest
coverage for \ecm near 30~\gev and the $Z^0$ peak.
Data from other experiments are consistent and yield the same
conclusions, but are omitted for clarity.
Strong scaling violations are evident,
both at low $x_p$ due to the pion mass and at high $x_p$ as expected
from the running of the strong coupling strength $\alpha_s$.
Also shown are the predictions of the JETSET model at these three
energies,
using default parameter values.
JETSET provides a good description of all three data sets for
$x_p\! >\!0.2$,
and hence describes the high-$x_p$ scaling violation well.
The other two models also reproduce this \ecm dependence,
though they do not describe the spectrum well at any energy.

The middle plot in Fig.~\ref{fig:xsqqscal} shows a similar test for
the \Kpm cross section.
Here we show the UCLA model predictions, 
as they describe our results best at high $x_p$.
The different flavor composition of the three samples is important for
\Kpm and modifies the expected scaling violation.
Kaons from $\bbbar$ events, which are absent from our data, 
contribute strongly to the TASSO cross section in the 0.1--0.3 region,
but little at higher $x_p$.
Since the cross sections are normalized per event, 
the expected scaling violation is reduced relative to that in the
\pipm cross sections in the 0.1--0.3 range, and increased at higher $x_p$.
At the $Z^0$ energy, the relative production of up- and down-type quarks
is quite different,
and the combination of more \Kpm from \bbbar and \ssbar events and
fewer from \ccbar events pushes the simulated high-$x_p$ cross section
up to nearly the same level as for the TASSO energy.

The flavor dependence has been shown~\cite{pikpsld,pikpdelphi} to be
accurately modeled at the $Z^0$ energy to the level of about 10\%.
The UCLA model describes the shape of the SLD cross section at high
$x_p$ well, 
but is too low by about 15\%.
The other models also predict about 15\% more scaling violation than
is observed.
However,
it is difficult to draw any conclusion in light of
the flavor dependence.

For protons, shown in the bottom plot in Fig.~\ref{fig:xsqqscal},
we compare with the JETSET model in which we have changed one
parameter value, 
the diquark production probability $P_{diqrk}$, from 0.1 to 0.085.
This provides a good description of the SLD and TASSO data at all $x_p$,  
although the latter are sparse at high $x_p$.
The simulated high-$x_p$ scaling violation between 10.54 and
34~\gev is similar to that for \pipm,
but that between 34 and 91~\gev is slightly larger since fast protons are
expected to be produced predominantly in \uubar and \ddbar events.
The prediction for 10.54~\gev is consistent with the \babar\ data for
$x_p$ below 0.07, but then rises well above the data, exceeding it by
as much as a factor of 3 at $x_p \!=\! 0.8$.
We see similar behavior for JETSET with default parameter values,
HERWIG, and UCLA.
Thus none of these models predicts the correct scaling properties for
protons, 
even though they describe the properties of pions well.

\subsection{Tests of MLLA QCD}
\label{sec:resqcd}

\begin{figure}[tbp]
 \begin{center} 
  \includegraphics[width=\hsize]{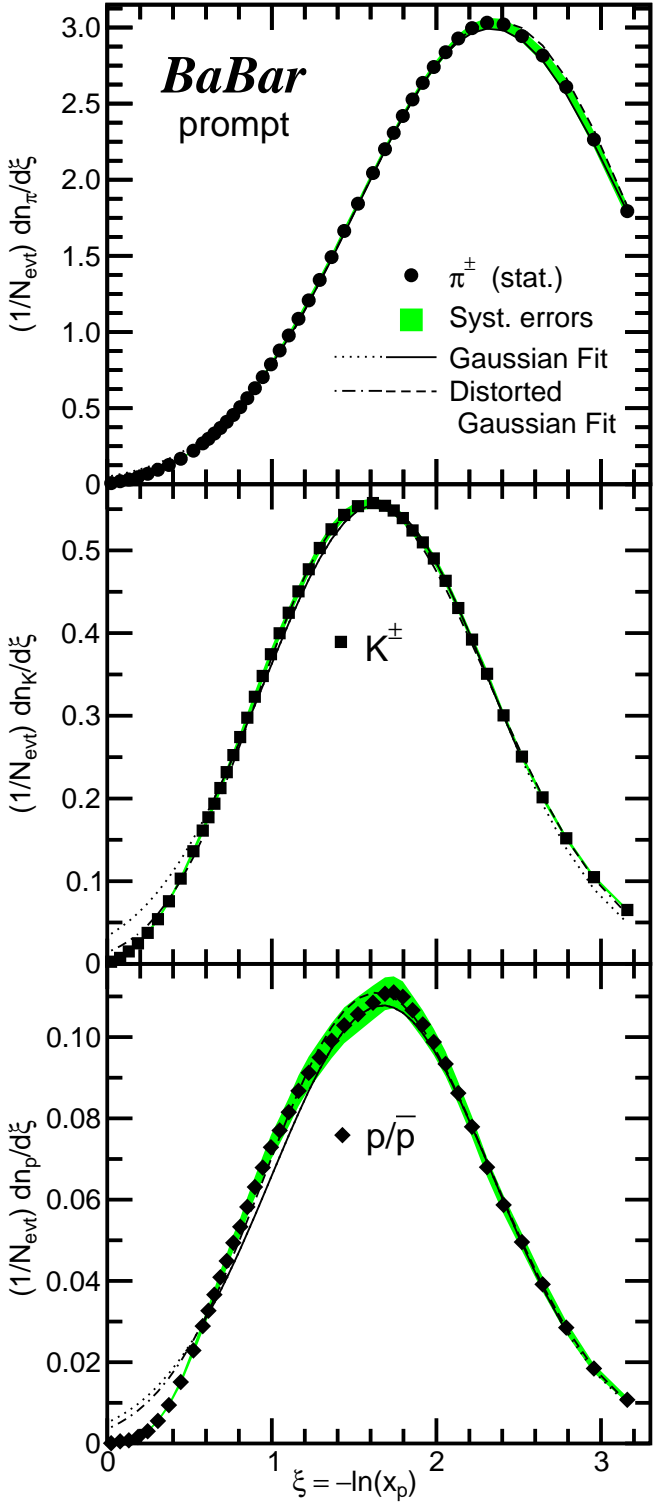}
 \end{center}
 \vspace{-0.6cm}
 \caption{ \label{fig:xifitqq}
   Differential cross sections in $\xi$ for prompt \pipm (top), 
   \Kpm (middle), and \ppbar (bottom).
   The error bars are statistical only,
   and the gray bands represent the systematic uncertainties, which are
   strongly correlated from point to point.
   Also shown are the results of the Gaussian (solid lines) and 
   distorted Gaussian (dashed lines) fits described in the text
   over their maximum ranges (see Table~\ref{tab:xifits});
   they are continued as dotted and dot-dashed lines, respectively,
   outside those ranges.
   }
\end{figure}

We test the predictions of QCD in the modified leading logarithm
approximation (MLLA)~\cite{mlla}, 
combined with the ansatz of local parton-hadron duality (LPHD)~\cite{mlla}, 
using our cross sections in the variable $\xi \!=\! -\ln(x_p)$.
Figure~\ref{fig:xifitqq} shows the $\xi$ distributions for prompt
particles;  the conventional distributions are similar in shape.
The error bars are statistical.
Because of their strong correlations,
the systematic uncertainties are shown as bands.
The normalization uncertainty is not included, as it does not affect
the shapes.

This representation emphasizes the low-\pstar (large $\xi$) region 
and most of each spectrum is visible on a linear vertical scale.
The spectra exhibit slow rises from zero at $\xi \!=\! 0$ 
(the beam momentum)
and the ``humpbacked plateau" predicted by MLLA$+$LPHD.
The MLLA$+$LPHD hypothesis also predicts that Gaussian functions
should describe these spectra over ranges of
$\pm$0.5--1 units about the peak position $\xi^*$,
and that slightly distorted Gaussian functions should fit the data
over substantially wider ranges.
Furthermore, $\xi^*$ should decrease exponentially with increasing
hadron mass at a given \ecm, 
and increase logarithmically with \ecm for a given hadron type.

Following convention, we first estimate $\xi^*$
by fitting Gaussian functions over a set of $\xi$ ranges
each about one unit wide and centered within one bin of the peak.
Given our binning, 
we consider nine such ranges for \pipm, and four for \Kpm and \ppbar.
We average the mean values and statistical and systematic
uncertainties of the set of fits,
and include the RMS deviation among the means as an additional
systematic uncertainty.
The fits use the full systematic error matrix and all have acceptable
$\chi^2$;
the resulting $\xi^*$ values are listed in Table~\ref{tab:xistar}
with their total uncertainties.
The statistical uncertainties are negligible,
and the RMS uncertainties are small except for the \pipm,
where they are about half the other systematics.
Adding more ranges to any set of fits has little effect on the results.

\begin{table}[tbp]
  \caption{
   Peak positions $\xi^*$, determined as described in the text,
   with total uncertainties, which are dominated by systematic terms.
    }
  \begin{center}
    \begin{tabular}{cccc}\hline\hline
                \multicolumn{4}{c}{  }                             \\[-0.3cm]
             &      \pipm      &       \Kpm      &      \ppbar     \\[0.1cm]
\hline\hline
                \multicolumn{4}{c}{  }                             \\[-0.3cm]
   Prompt    & 2.337$\pm$0.009 & 1.622$\pm$0.006 & 1.647$\pm$0.019 \\
Conventional & 2.353$\pm$0.009 & 1.622$\pm$0.006 & 1.604$\pm$0.013 \\ 
\hline\hline
    \end{tabular}
  \end{center}
\label{tab:xistar}
\end{table}

To test the prediction regarding the Gaussian shape,
we first
find the largest range centered near $\xi^*$ over which the Gaussian
fit is acceptable, 
i.e., yields a $\chi^2$ with a confidence level exceeding 0.01. 
We then extend the fit range to either lower or higher values, if
possible,
to find a maximum range over which this function gives an acceptable
fit. 
These ranges are listed in Table~\ref{tab:xifits} and the fits are
shown as the lines on Fig.~\ref{fig:xifitqq}.
In each case we obtain a good fit over a range more than one unit wide,
consistent with the prediction.
For \Kpm and \ppbar the maximum ranges are centered near the peak and
span nearly two units in width.
The maximum range for pions extends to the end of our coverage, 
which is just under one unit above the peak.
It extends more than 1.5 units below the peak, 
but data at higher $\xi$ might constrain this more tightly.

\begin{table}[tbp]
  \caption{
   Maximum ranges over which we obtain good fits using a
   simple Gaussian function and a distorted Gaussian function that
   includes skewness and kurtosis terms.
   }
  \begin{center}
    \begin{tabular}{ccccccccccc}\hline\hline
          \multicolumn{11}{c}{  }                                  \\[-0.3cm]
        & \multicolumn{5}{c}{Prompt}
        & \multicolumn{5}{c}{Conventional} \\ 
 Hadron && Gaussian   && Distorted  &&& Gaussian   && Distorted  & \\
  \hline\hline
          \multicolumn{11}{c}{  }                                  \\[-0.3cm]
 \pipm  && 0.92--3.27 && 0.22--3.27 &&& 0.87--3.27 && 0.67--3.27 & \\
 \Kpm   && 0.63--2.58 && 0.34--3.05 &&& 0.63--2.58 && 0.34--3.05 & \\
 \ppbar && 0.56--3.27 && 0.48--3.27 &&& 0.71--2.58 && 0.48--3.27 & \\
   \hline\hline
    \end{tabular}
  \end{center}
\label{tab:xifits}
\end{table}

Next, we add approximate skewness ($s$) and kurtosis ($\kappa$) terms
to our fitting function, following Ref.~\cite{distgauss}:
\[
G^\prime(\xi) =
\frac{N}{\sigma \sqrt{2\pi}}
\exp\left(\frac{\kappa}{8} + \frac{s\delta}{2}
- \frac{(2 \!+\! \kappa)\delta^{2}}{4}
+ \frac{s\delta^{3}}{6}
+ \frac{\kappa\delta^{4}}{24} \right) ,  \nonumber
\]
where $\delta\! =\! (\xi-\xi^*)/\sigma$, 
and $\sigma$ is the standard deviation.
We repeat the exercise of finding the maximum $\xi$ range
for which a fit of this function is acceptable.
The results are listed in Table~\ref{tab:xifits} and shown as the dashed
lines on Fig.~\ref{fig:xifitqq}.
The prompt (conventional) \pipm range can be extended to substantially
(somewhat) lower $\xi$ values,
consistent with the prediction.
However, 
the fitted skewness and kurtosis values increase rapidly as the range is
extended,
to $-$0.37 ($-$0.11) and $-$0.43 ($-$0.30), respectively, at the
widest range,
and it is unknown how additional data on the high side of the peak
might affect the fits.
The \Kpm and \ppbar ranges can be extended somewhat in
both directions with similar $s$ and $\kappa$ values.
Given our relatively low \ecm and hence narrow $\xi$ range,
this should also be considered consistent with the prediction.

\begin{figure}[tbp]
 \begin{center} 
  \includegraphics[width=\hsize]{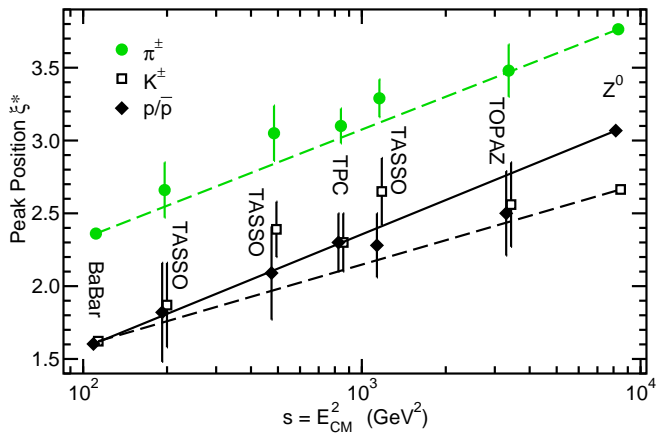}
 \end{center}
 \caption{ \label{fig:xistare} 
   Peak positions $\xi^*$ vs.\ CM energy for pions (circles), 
   kaons (squares) and protons (diamonds) on a logarithmic horizontal
   scale.
   The lines join our points with those from the averages of
   the $Z^0$ experiments.
   }
\end{figure}

We find $\xi^*_\pi$ to be 0.8 units higher than $\xi^*_K$, 
consistent with the predicted decrease with hadron mass,
but $\xi^*_p$ is not lower than $\xi^*_K$.
This is similar to the behavior observed at higher energies
where mesons and baryons appear to follow different trajectories,
but measurements for more particles at our \ecm would be needed
to draw firm conclusions.
In Fig.~\ref{fig:xistare} we show a compilation of $\xi^*$ measurements
for \pipm, \Kpm and \ppbar as a function of CM energy.
Our precise values and those from the $Z^0$ provide strong constraints
on the trajectories,
and the lines on the plot simply join the points at these two energies.
The other data are consistent with the lines,
and hence with the predicted energy dependence,
but more precise data at other energies are needed to test the form of
the increase.
The slopes of the lines for pions and protons are similar,
but that for the kaons is quite different.
This could be due to the changing flavor composition.

\subsection{Average Multiplicities, Ratios and Fractions}
\label{sec:restot}

To estimate the average numbers of pions, kaons and protons produced
per event,
we integrate the differential cross sections over the measured \pstar
range, 
and correct for the unmeasured parts of the spectra.
The integrals take all systematic uncertainties and their correlations
into account,
and are listed in the second column of Table~\ref{tab:ratecont}.
The uncertainties are dominated by the normalization and fully
correlated tracking systematics;
there are also substantial contributions to the conventional \pipm and
\ppbar results from the \KS and strange baryon cross sections.

\begin{table*}[tbp]
\caption{
   Integrated measured cross sections,
   the fractional coverage estimated as described in the text,
   and the fully corrected \eeqqb event multiplicities from this
   measurement.
   The first error on the yield is experimental, 
   dominated by systematics, 
   and the second is from the uncertainty on the coverage.
   The other columns show previous results from the CLEO~\cite{cleohad}
   and ARGUS~\cite{argus89} experiments, 
   and values predicted by the models described in the text.
   }
  \begin{center}
    \begin{tabular}{|l@{}c||l@{$\pm$}l|c||l@{$\pm$}l@{$\pm$}l|c|c||c|c|c|}
   \hline
&          & \multicolumn{2}{c|}{ }
          && \multicolumn{8}{c|}{  } \\[-0.3cm]
&          & \multicolumn{2}{c|}{Measured}
          && \multicolumn{8}{c|}{Yield per \qqbar Event} \\
   \cline{6-13}
&          & \multicolumn{2}{c|}{ }
          && \multicolumn{3}{c|}{ } &&&&& \\[-0.25cm]
             \multicolumn{2}{|c||}{Particle}
           & \multicolumn{2}{c|}{Integral}        & Coverage        
           & \multicolumn{3}{c|}{\babar}          &  CLEO    &  ARGUS
           & JETSET & UCLA & HERWIG    \\
   \hline
&          & \multicolumn{2}{c|}{ } && \multicolumn{3}{c|}{ } &&&&& \\[-0.3cm]
& \emppm   & 5.51  & 0.08  & 0.876$\pm$0.018 & 6.29  & 0.09  & 0.13
           &               &                 & 5.84  & 5.88  & 5.73 \\
Prompt
&  \pipm   & 5.36  & 0.08  & 0.884$\pm$0.019 & 6.07  & 0.09  & 0.13
           &               & 5.694$\pm$0.108 & 5.59  & 5.62  & 5.49 \\
&   \Kpm   & 0.946 & 0.012 & 0.973$\pm$0.016 & 0.972 & 0.012 & 0.016
           &               & 0.888$\pm$0.030 & 1.01  & 1.02  & 1.01 \\
& \ppbar   & 0.182 & 0.006 & 0.984$\pm$0.007 & 0.185 & 0.006 & 0.001
           &               & 0.212$\pm$0.017 & 0.28  & 0.14  & 0.31 \\ 
  \hline
&          & \multicolumn{2}{c|}{ } && \multicolumn{3}{c|}{ } &&&&& \\[-0.3cm]
& \emppm   & 6.15  & 0.10  & 0.867$\pm$0.019 & 7.09  & 0.11  & 0.16
           &               &                 & 6.58  & 6.60  & 6.55 \\
Conventional
&  \pipm   & 6.00  & 0.10  & 0.874$\pm$0.020 & 6.87  & 0.11  & 0.16
           &  8.3$\pm$0.4  &  6.38$\pm$0.12  & 6.33  & 6.34  & 6.31 \\
&   \Kpm   & 0.946 & 0.012 & 0.973$\pm$0.016 & 0.972 & 0.012 & 0.016
           &  1.3$\pm$0.2  & 0.888$\pm$0.030 & 1.01  & 1.02  & 1.01 \\
& \ppbar   & 0.261 & 0.008 & 0.984$\pm$0.008 & 0.265 & 0.008 & 0.002
           & 0.40$\pm$0.06 & 0.271$\pm$0.018 & 0.37  & 0.20  & 0.46 \\
  \hline
    \end{tabular}
  \end{center}
\label{tab:ratecont}
\end{table*}

From Fig.~\ref{fig:xifitqq}, it is clear that the coverage, 
i.e. the fraction of the spectrum covered by our measurement, 
is over 95\% for \Kpm and \ppbar.
However, it is smaller for \pipm, and in no case is it clear a priori
how to account for this reliably.
We consider four estimates of our coverage, 
one from each of the three hadronization models
and one from an ensemble of distorted Gaussian fits.
We consider fits over all ranges that include the ten highest-$\xi$
points and give an acceptable $\chi^2$ calculated from only the bins
above the peak plus the five bins just below the peak.
The average of these four coverage values is given in the third column of
Table~\ref{tab:ratecont},
with an uncertainty that corresponds to their RMS deviation.
The spread among the fits is smaller than this,
as are variations obtained by running any simulation with 
different parameter values.
We divide each measured integral by the corresponding coverage to
obtain the average event multiplicity listed in column four of 
Table~\ref{tab:ratecont}.

Previous results from CLEO at 10.49~\gev~\cite{cleohad}
and ARGUS at 9.98~\gev are also listed in Table~\ref{tab:ratecont},
as are the predictions of the three hadronization models.
Our prompt (conventional) \pipm rate is 7\% (8\%) and 2.0$\sigma$
(2.2$\sigma$) higher than the ARGUS rate.
A difference of this size is expected from the \ecm difference.
Our \Kpm and \ppbar rates are also slightly higher than the ARGUS rates.
The CLEO rates are substantially higher than ours, 
but their uncertainties are large.
With default parameter values, 
all three models give conventional \pipm rates close to the ARGUS
value and 8-9\% below ours,
even though the simulations are run at our \ecm.
The models predict \Kpm rates that are slightly too high,
and widely varying \ppbar rates, 
none of which is consistent with the data.
The total charged hadron rates from ARGUS and CLEO are among the main
inputs to the tuning of these models.

\begin{figure}[tbp]
 \begin{center}
  \includegraphics[width=0.99\hsize]{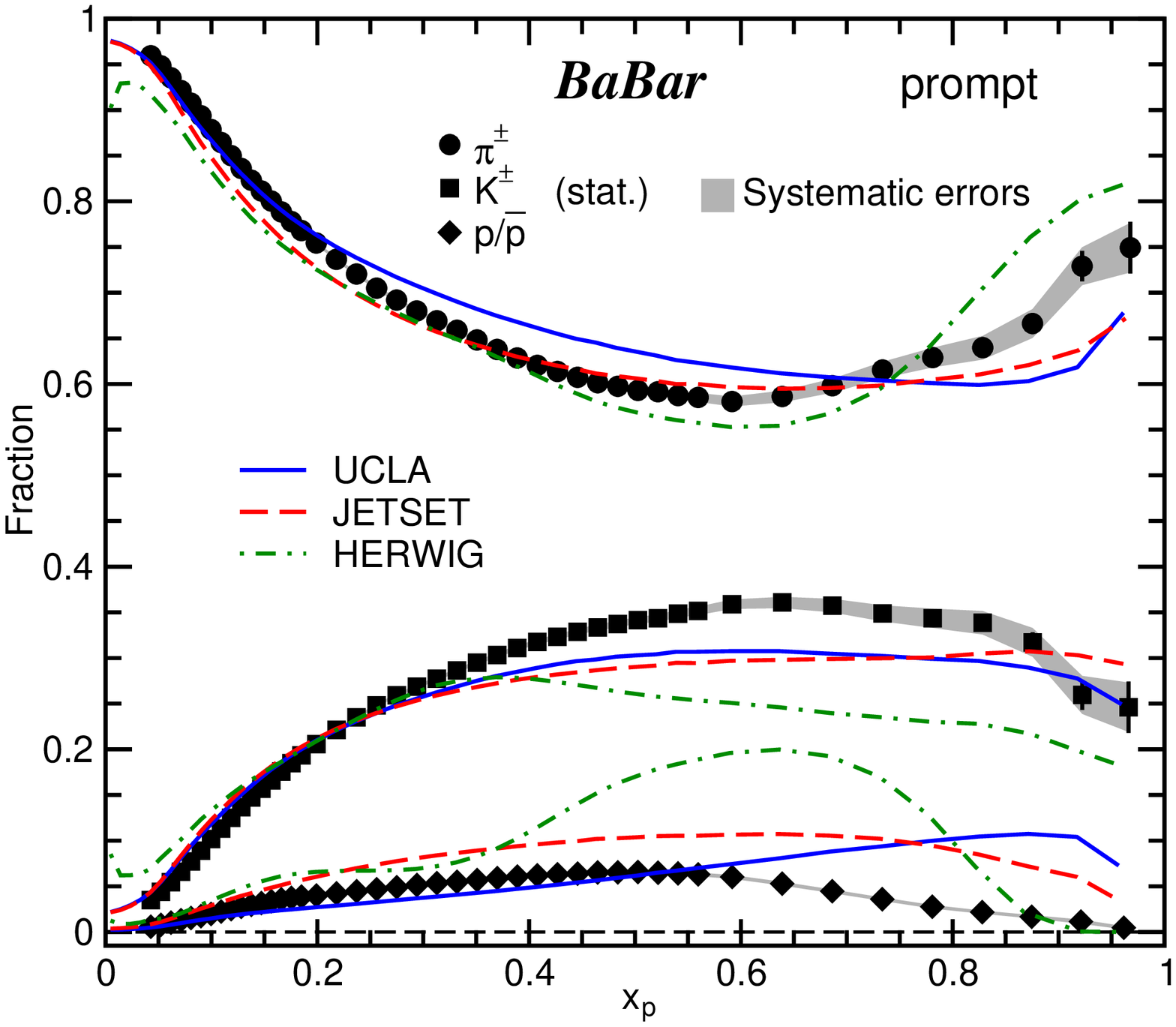}
  \includegraphics[width=0.99\hsize]{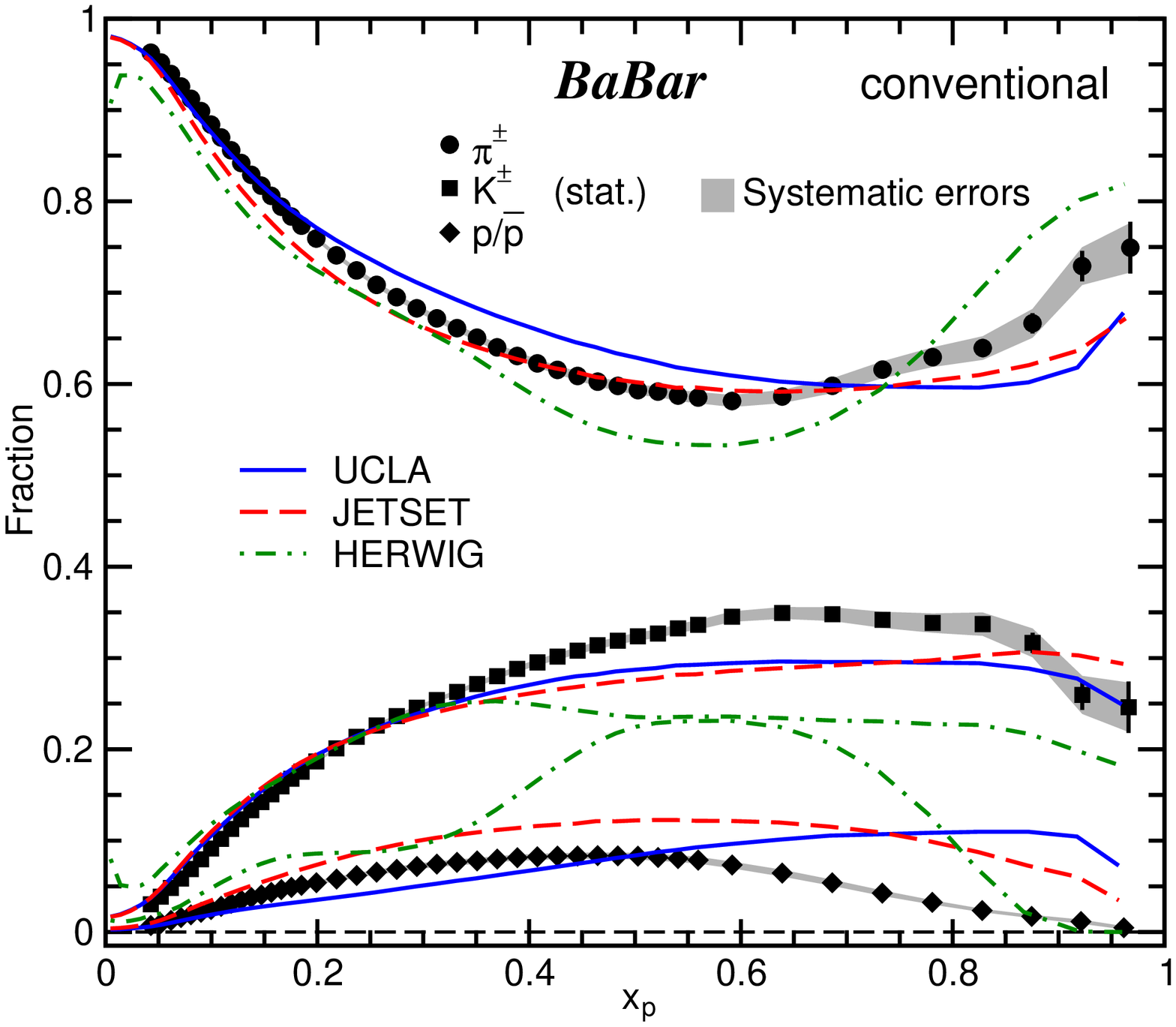}
 \end{center}
 \caption{ \label{fig:fraxqq} 
   Prompt (top) and conventional (bottom) \pipm, \Kpm, and \ppbar
   fractions.
   The error bars are statistical only,
   and the gray bands represent the systematic uncertainties, which are
   strongly correlated from point to point.
   Also shown are the predictions of the three hadronization models.
   }
\end{figure}

From our cross sections we can derive production ratios for pairs of
hadrons,
in which many of the systematic uncertainties cancel at least partly.
The remaining uncertainties are dominated by particle identification
systematics.
Previous experiments have presented this information in the form of
the fractions of all charged hadrons that are pions, kaons and
protons, $f_\pi$, $f_K$ and $f_p$.
We show our fractions for prompt and conventional hadrons in
Fig.~\ref{fig:fraxqq},
and tabulate them in the supplementary material~\cite{supmat}.
The prompt and conventional fractions are quite similar,
and converge at high $x_p$.
Strange hadron decay products cause the 
conventional $f_\pi$ and $f_p$ to be larger than their prompt
counterparts at low $x_p$,
with $f_K$ correspondingly smaller.

The dominance of pions at low $x_p$ is expected due to their lower
mass and the contributions from decays of heavier hadrons.
The plateau values of $f_K$ and $f_p$ near $x_p \!=\! 0.6$,
of about 0.35 and 0.08, respectively,
might reflect the intrinsic relative production of strange particles
and baryons in the hadronization process.
The decrease of $f_p$ at high $x_p$ might be kinematic -- 
a proton must be produced along with an antibaryon, 
and the mass of the pair is a large fraction of \ecm /2.
The \Kpm from \ccbar events are also kinematically limited, 
whereas those from \ssbar and \uubar events become more important as
$x_p$ increases, until perhaps the very highest-$x_p$ bins.

The predictions of the three models are also shown,
and do not describe the data well.
JETSET and UCLA provide reasonable qualitative descriptions
but underpredict $f_K$ and overpredict $f_p$ at high $x_p$.
In particular $f_p$ does not decrease early or quickly enough.
HERWIG's description of $f_p$ is poor, 
and this affects $f_\pi$ and $f_K$, 
which might otherwise be described reasonably well.

\section{Summary}
\label{sec:summary}

We present measurements of the differential production cross sections 
for charged pions, kaons, and protons in \epem annihilations at
$\ecm = 10.54~\gev$,
both excluding (prompt) and including (conventional) decay products
of \KS mesons and weakly decaying strange baryons.
The measurements cover the CM momentum (\pstar) range from 0.2~\gevc
to the beam momentum.
Comparing with previous measurements at the nearby \ecm of 9.98~\gev,
we find consistency for \pstar in the 1--3~\gevc range,
and evidence for scaling violations below 1~\gevc.

These data can be used to test and tune models of the hadronization
process.
We find that the JETSET, UCLA, and HERWIG models, 
which were tuned to previous data between 9.98 and 35~\gev,
reproduce the \pipm and \Kpm spectra to within 15\% over most of
the \pstar range,
but do not describe their shapes in detail.
All three models provide poor descriptions of the \ppbar spectra.
Comparing the same models with data at higher \ecm,
we find that they reproduce the high-\pstar scaling properties
of the \pipm cross section to within a few percent 
and the \Kpm spectrum to within 15\%, 
but predict about twice the scaling violation observed for \ppbar.

The shape of the $\xi \!=\! -\ln(x_p)$ spectrum predicted by MLLA QCD
is consistent with our data in all cases,
and the peak positions $\xi^*$ are lower for \Kpm than \pipm,
as predicted.
However, the $\xi^*$ for \ppbar are not lower than those for \Kpm.
This is consistent with the behavior observed at higher \ecm,
where the predicted mass dependence holds for mesons and baryons
separately, but not together.
The predicted \ecm dependence is consistent with the world's data, 
with the slopes being similar for \pipm and \ppbar; 
the \Kpm slope is lower, 
perhaps due to the changing flavor composition with increasing \ecm.

We integrate over the measured \pstar ranges,
and extrapolate into the unmeasured regions,
to measure a total of
$6.07\pm0.16$, $0.97\pm0.02$, and $0.19\pm0.01$ prompt
($6.87\pm0.19$, $0.97\pm0.02$, and $0.27\pm0.01$ conventional)
\pipm, \Kpm and \ppbar, respectively, per hadronic event.
We also provide hadron fractions, in which many of the systematic
uncertainties cancel.
These measurements are also consistent with previous results and
provide additional information that can be used to test models.

\section*{Acknowledgements}
We are grateful for the 
extraordinary contributions of our \pep2\ colleagues in
achieving the excellent luminosity and machine conditions
that have made this work possible.
The success of this project also relies critically on the 
expertise and dedication of the computing organizations that 
support \babar.
The collaborating institutions wish to thank 
SLAC for its support and the kind hospitality extended to them. 
This work is supported by the
US Department of Energy
and National Science Foundation, the
Natural Sciences and Engineering Research Council (Canada),
the Commissariat \`a l'Energie Atomique and
Institut National de Physique Nucl\'eaire et de Physique des Particules
(France), the
Bundesministerium f\"ur Bildung und Forschung and
Deutsche Forschungsgemeinschaft
(Germany), the
Istituto Nazionale di Fisica Nucleare (Italy),
the Foundation for Fundamental Research on Matter (The Netherlands),
the Research Council of Norway, the
Ministry of Education and Science of the Russian Federation, 
Ministerio de Econom\'{\i}a y Competitividad (Spain), and the
Science and Technology Facilities Council (United Kingdom).
Individuals have received support from 
the Marie-Curie IEF program (European Union) and the A. P. Sloan Foundation (USA).



\begin{thebibliography}{99} 

\bibitem{ert}
See, e.g., R.K.~Ellis, D.A.~Ross and A.E.~Terrano, 
Nucl.\ Phys.\ B {\bf 178}, 421 (1981).

\bibitem{moretti}
See, e.g., S.~Moretti, Phys.\ Lett.\ B {\bf 420}, 367 (1998).

\bibitem{mlla}
Y.I.~Azimov, Y.L.~Dokshitzer, V.A.~Khoze, and S.I.~Troian,
Z.\ Phys.\ C {\bf 27}, 65 (1985).

\bibitem{nlla}
G.~Marchesini and B.R.~Webber, Nucl.\ Phys.\ B {\bf 238}, 1 (1984).

\bibitem{kkp}
S.~Albino, B.A.~Kniehl, and G.~Kramer, 
Nucl.\ Phys.\ B {\bf 803}, 42 (2008);
D.~de~Florian, R.~Sassot, and M.~Stratmann,
Phys.\ Rev.\ D {\bf 75}, 114010 (2007);
M.~Hirai, S.~Kumano, T.H.~ Nagai, and K.~Sudoh,
Phys.\ Rev.\ D {\bf 75}, 094009 (2007).

\bibitem{bohrer}
See, e.g., A.~B\"ohrer, 
Phys.\ Rep.\ {\bf 291}, 107 (1997);
G.D.~Lafferty, P.J.~Reeves and M.R.~Whalley, 
J. Phys. G{\bf 21}, A1 (1995);
D.H.~Saxon, 
in {\it High Energy Electron-Positron Physics Vol. 1}, 
Eds. A.~Ali and P.~S\"oding, World Scientific (1988) 539.

\bibitem{herwig}
G.~Corcella et al., 
JHEP {\bf 0101}, 010 (2001);
G.~Marchesini et al., 
Comput.\ Phys.\ Commun.\ {\bf 67}, 465 (1992).

\bibitem{jetset}
T.~Sj\"ostrand,
Comput.\ Phys.\ Commun.\ {\bf 82}, 74 (1994).

\bibitem{ucla}
S.~Chun and C.~Buchanan,
Phys.\ Rep.\  {\bf 292}, 239 (1998).

\bibitem{delphias}
P.~Abreu et al. (DELPHI Collaboration), 
Phys.\ Lett.\  B {\bf 398}, 194 (1997).

\bibitem{argus89}
H.~Albrecht et al. (ARGUS Collaboration),
Z.\ Phys.\ C {\bf 44}, 547 (1989).

\bibitem{pikptpc}
H.~Aihara et al. (TPC-2Gamma Collaboration), 
Phys.\ Rev.\ Lett.\  {\bf 61}, 1263 (1988).

\bibitem{pikptasso}
W.~Braunschweig et al. (TASSO Collaboration),
Z.\ Phys.\ C {\bf 42}, 189 (1989).

\bibitem{pikptopaz}
R.~Itoh et al. (TOPAZ Collaboration), 
Phys.\ Lett.\ B {\bf 345}, 335 (1995).

\bibitem{pikpdelphi}
P.~Abreu et al. (DELPHI Collaboration), 
Eur.\ Phys.\ J.\ C {\bf 5}, 585 (1998).

\bibitem{pikpopal}
R.~Akers et al. (OPAL Collaboration), 
Z.\ Phys.\ C {\bf 63}, 181 (1994).

\bibitem{pikpaleph}
D.~Buskulic et al. (ALEPH Collaboration), 
Z.\ Phys.\ C {\bf 66}, 355 (1995).

\bibitem{pikpsld}
K.~Abe et al. (SLD Collaboration), 
Phys.\ Rev.\ D {\bf 59}, 052001 (1999);
Phys.\ Rev.\ D {\bf 69}, 072003 (2004).

\bibitem{pikpdelpiw}
P.~Abreu et al. (DELPHI Collaboration), 
Eur.\ Phys.\ J.\ C {\bf 18} (2000) 203;
 [Erratum-ibid.\ C {\bf 25} (2002) 493].

\bibitem{belle13}
M.~Leitgab, et al. (Belle Collaboration),
Phys.\ Rev.\ Lett.\  {\bf 111} (2013) 062002.

\bibitem{lpsld} 
K.~Abe et al. (SLD Collaboration),
Phys.\ Rev.\ Lett.\  {\bf 78} (1997) 3442;
    [Erratum-ibid.\  {\bf 79} (1997) 959].

\bibitem{lpopal} 
G.~Abbiendi et al. (OPAL Collaboration),
Eur.\ Phys.\ J.\ C {\bf 16}, 407 (2000).

\bibitem{babarNIM}
B.~Aubert et al. (\babar\ Collaboration),
Nucl.\ Instr.\ Methods A {\bf 479}, 1 (2002);
arXiv:1305.3560 [hep-ex],
submitted to Nucl.\ Instr.\ Methods.

\bibitem{dircNIM}
I.~Adam et al. (\babar\ DIRC Collaboration), 
Nucl.\ Instr.\ Methods A {\bf 538}, 281 (2005).

\bibitem{foxwolfram}
G.C.~Fox and S.~Wolfram,
Phys.\ Rev.\ Lett.\ {\bf 41}, 1581 (1978);
Nucl.\ Phys.\ B {\bf 149}, 413 (1979).

\bibitem{thrust}
S.~Brandt et al., Phys.\ Lett.\ {\bf 12}, 57 (1964);
E.~Farhi, Phys.\ Rev.\ Lett.\ {\bf 39}, 1587 (1977).

\bibitem{geant4}
S.~Agostinelli et al. (GEANT4 Collaboration), 
Nucl.\ Instr.\ Methods A {\bf 506}, 250 (2003).

\bibitem{koralb}
S.~Jadach and Z.~Was,
Comput.\ Phys.\ Commun.\  {\bf 64}, 267 (1991).

\bibitem{pdg}
J.~Beringer, et al. (Particle Data Group), 
Phys.\ Rev.\ D {\bf 86}, 010001 (2012).

\bibitem{bhwide}
S.~Jadach, W.~Placzek, and B.F.L.~Ward,
Phys.\ Lett.\ B {\bf 390}, 298 (1997).

\bibitem{afqed}
H.~Czyz and J.H.~K\"uhn, 
Eur.\ Phys.\ J.\ C {\bf 18}, 497 (2001).

\bibitem{gamgam}
H.~Paar and M.~Sivertz, 
(unpublished), based on
V.M.~Budnev, et al., Phys.\ Rep.\ {\bf 15}, 181 (1975).

\bibitem{cleohad}
S.~Behrends et al. (CLEO Collaboration), 
Phys.\ Rev.\ D {\bf 31}, 2161 (1985).

\bibitem{argusk0lam}
H.~Albrecht et al. (ARGUS Collaboration), 
Z.\ Phys.\ C {\bf 62}, 371 (1994).

\bibitem{brandon}
B.L.~Hartfiel, Ph.D.\ Thesis, Univ. of California-Los Angeles (2004); 
SLAC-R-823 (unpublished).

\bibitem{tkeff}
T.~Allmendinger et al., 
Nucl.\ Instr.\ Methods Phys.\ Res.\ Sect.\ A {\bf 704} (2013) 44.

\bibitem{schrist}
S.~Christ, Ph.D.\ Thesis, Universit\"{a}t Rostock (2003) (unpublished).

\bibitem{bbrLcmass}
B.~Aubert, et al. (\babar\ Collaboration), 
Phys.\ Rev.\ D {\bf 72}, 052006 (2005).

\bibitem{supmat}
Tables of charged hadron fractions, uncertainties and correlation
matrices are available through Phys.\ Rev.

\bibitem{cleocharm}
M.~Artuso et al. (CLEO Collaboration),
Phys.\ Rev.\ D {\bf 70}, 112001 (2004).

\bibitem{bellecharm}
R.~Seuster et al. (Belle Collaboration),
Phys.\ Rev.\ D {\bf 73}, 032002 (2006).

\bibitem{bbrLcspect}
B.~Aubert, et al. (\babar\ Collaboration),
Phys.\ Rev.\ D {\bf 75}, 012003 (2007).

\bibitem{distgauss}
Y.L.~Dokshitzer, V.A.~Khoze, and S.I.~Troian,
Int.\ J.\ Mod.\ Phys.\ A {\bf 7}, 1875 (1992);
C.P.~Fong and B.R.~Webber,
Phys.\ Lett.\ B {\bf 229}, 289 (1989).



\end{thebibliography}
\end{document}